\documentclass[12pt]{article}
\usepackage[margin=1in]{geometry}
\usepackage{setspace}
\usepackage{natbib}
\usepackage{hyperref} 
\usepackage{enumitem}
\usepackage{titlesec}

\usepackage{mathrsfs}
\usepackage{amsmath, amssymb, amsthm}
\allowdisplaybreaks[4]
\usepackage{CJK}
\usepackage{bm}
\usepackage{caption}
\usepackage{graphicx, subfig}
\usepackage{array}
\usepackage{booktabs}
\usepackage{appendix}
\usepackage{multirow} 

\usepackage{color,xcolor}
\usepackage{algorithm, algorithmic}

\usepackage{authblk}  % 用于编号和机构分配

\usepackage[normalem]{ulem}

\usepackage[normalem]{ulem}
\usepackage{cancel}

\newtheorem{assumption}{Assumption}
\newtheorem{subassumption}{Assumption}[assumption]
\newtheorem{theorem}{Theorem}

\newtheorem{assumptioninner}{Assumption}[assumption]

\newenvironment{assumptionprime}
  {
   \assumptioninner}
  {\endassumptioninner}

\title{\bf Synthetic Control Method with Mixed Frequency Data} %% Article title

\author[1]{Lu Zhang}
\author[1]{Shijin Gong}
\author[2,1]{Xinyu Zhang\thanks{Corresponding author. Email: \texttt{xinyu@amss.ac.cn}}}

\affil[1]{School of Management, University of Science and Technology of China, Hefei, China}

\affil[2]{Academy of Mathematics and Systems Science, Chinese Academy of Sciences, Beijing, China}

\date{}

\begin{document}

\maketitle

%% Abstract
\begin{abstract}
%% Text of abstract
Mixed-frequency data, where variables are observed at different temporal resolutions, commonly occur in economic and financial studies. Classical synthetic control methods (SCM) are ill-suited for such data, often necessitating aggregation or prefiltering that may discard valuable information. This paper proposes a novel Mixed-Frequency Synthetic Control Method (MF-SCM) to integrate mixed-frequency data into the synthetic control framework effectively.
We develop a flexible estimation procedure to construct synthetic control weights under mixed-frequency settings and establish the theoretical properties of the MF-SCM estimator. Specifically, we first prove that the estimator achieves asymptotic optimality, in the sense that it achieves the lowest possible squared prediction error among all potential treatment effect estimators from averaging outcomes of control units. We then derive the asymptotic distribution of the average treatment effect (ATE) estimator using projection theory and construct confidence intervals for the ATE estimator. The method's effectiveness is demonstrated through numerical simulations and two empirical applications concerning the 2017 Tax Cuts and jobs Act in US and air pollution alerts.\\
\end{abstract}

%% Keywords
\vspace{1em}

%% Add \usepackage{lineno} before \begin{document} and uncomment 
%% following line to enable line numbers
%% \linenumbers

%% main text
%%

%% Use \section commands to start a section

\setcounter{equation}{0}

\numberwithin{equation}{section}
\section{Introduction}
Evaluating the causal effects of interventions or policies is a cornerstone of empirical research across economics, social sciences, and public policy.
The synthetic control method (SCM), proposed by \citet{abadie2003economic} and further developed by \citet{abadie2010synthetic}, has emerged as a fundamental tool for assessing causal effects using observational data.
The basic idea of SCM is to construct a synthetic control unit---a weighted combination of control units---that closely approximates the pre-treatment characteristics of the treated unit. By doing so, the post-treatment outcomes of the synthetic control can serve as counterfactual estimates for the treated unit's outcomes.
Since its inception, SCM has been recognized as one of the most significant innovations in policy evaluation methodologies, garnering widespread application across various domains (\citeauthor{athey2017state}, \citeyear{athey2017state}).
Despite its demonstrated effectiveness in numerous applications, the traditional implementation framework of SCM primarily relies on data sampled at a uniform frequency. This limitation reduces its applicability in contexts involving mixed-frequency data, such as quarterly economic indicators paired with monthly financial data or annual policy outcomes.

Mixed-frequency data, characterized by variables observed at different sampling frequencies, frequently arise in economic and financial studies.
The availability of data sampled at different frequencies poses a dilemma for a researcher working with time series data. On the one hand, high-frequency variables often contain rich and timely information. On the other hand, the researcher cannot use this high-frequency information directly if some of the variables are available at a lower frequency, because most time series regressions require data to be sampled at the same interval. The common solution in such cases is to prefilter the data so that the left-hand and right-hand variables are available at the same frequency. However, in the process, a lot of potentially useful information might be discarded, which can compromise the accuracy of subsequent inference and analysis.
To address these challenges, several methodologies have been developed within the field of econometrics, including bridge equations and mixed-data sampling (MIDAS) regression; see, for example, \citet{foroni2013survey} for a comprehensive survey. Bridge-equation approaches typically aggregate high-frequency variables to the low-frequency scale before estimation and are widely used in forecasting applications.
By contrast, MIDAS regression, introduced by \citet{ghysels2004midas}, allows researchers to model relationships between variables observed at different frequencies without first aggregating the high-frequency variables into low-frequency aggregates. These models provide a flexible framework for handling data sampled at different frequencies and enable direct forecasting of low-frequency variables (\citeauthor{ghysels2004midas}, \citeyear{ghysels2004midas}; \citeauthor{clements2008macroeconomic}, \citeyear{clements2008macroeconomic}).

Recent studies by \citet{sun2024temporal} have pointed out important considerations for applying SCM with high-frequency data. \citet{sun2024temporal} highlight that the large number of pre-treatment observations in high-frequency settings can amplify the risk of bias by overfitting to noise. Relatedly, \citet{abadie2022synthetic} caution about such bias from using disaggregated outcomes in SCM. A common strategy to mitigate this issue, as discussed in \citet{sun2024temporal}, is to aggregate high-frequency outcomes into a lower-frequency series (e.g., annual averages) and then estimate synthetic weights based on the aggregated data. \citet{sun2024temporal} further analyze the bias–variance trade-off associated with aggregating high-frequency outcomes and provide conditions under which temporal aggregation improves bias bounds, thereby motivating the importance of exploiting high-frequency information in SCM. 
Despite these insights, relatively limited attention has been paid to adapting SCM to mixed-frequency settings. This gap is significant given the increasing availability of mixed-frequency data and the need for causal analyses that incorporate both high-resolution and low-resolution information. 
This paper seeks to bridge this gap by proposing a novel Mixed-Frequency Synthetic Control Method (MF-SCM). The proposed approach extends the SCM framework to accommodate mixed-frequency data, enabling researchers to construct more accurate counterfactuals in settings where data are recorded at different intervals.

In this paper, we propose a flexible estimation procedure to obtain the synthetic control (SC) weight under mixed-frequency data. The proposed methodology addresses the challenges posed by differing sampling frequency across variables, enabling integration of such data into the synthetic control framework.
First, we detail the implementation steps of MF-SCM, providing a step-by-step guide for researchers to apply this approach in practice.
Next, we establish the MF-SCM estimator's asymptotic optimality, in the sense that it achieves the lowest possible squared prediction error among all potential treatment effect estimators from averaging outcomes of control units, as the number of pretreatment periods goes to infinity. This result underscores the effectiveness of our method in estimating treatment effects, even in complex data environments.
In the synthetic control literature, \citet{zhang2022asymptotic} and \citet{zhang2024asymptotic} also study the large sample properties of SC estimators. 
Additionally, the asymptotic optimality of the MF-SCM estimator, established in this paper, does not rely on the model structure. In other words, it does not need to assume the data generating process (DGP) of the potential outcomes. Our asymptotic optimality holds under a model-free setup, making it broadly applicable across diverse settings. This generality ensures that our methodology subsumes many existing models as special cases, highlighting its versatility and wide-ranging applicability in empirical research. Furthermore, we examine the theoretical distribution properties of the MF-SCM estimator and develop a comprehensive inference methodology tailored to the MF-SCM framework.

Our paper is also related to a broader class of panel data methods for treatment effect estimation. 
Methods such as generalized synthetic control, interactive fixed effects, matrix completion, and synthetic difference-in-differences typically require outcomes and covariates to be organized as a common unit-time panel before counterfactual outcomes can be estimated. In mixed-frequency applications, however, such a common panel structure is not directly available. 
The frequency-alignment idea underlying MF-SCM---including the latent-variable component for lower-frequency outcomes and the MIDAS-style component for higher-frequency outcomes---provides a possible preprocessing or modeling steps for constructing baseline-frequency representations of control outcomes prior to applying these related estimators. 
Thus, although this paper focuses on SCM, the proposed framework suggests a general route for adapting panel data treatment-effect methods to mixed-frequency environments. A formal theoretical analysis of these extensions is beyond the scope of this paper and is left for future research.

Our work connects with \citet{sun2024temporal}.  
Their paper provides a careful finite-sample analysis of the trade-off between aggregating high-frequency outcomes and using disaggregated series in SCM and gives conditions under which aggregation improves bias bounds. However, their analysis focuses on a single-frequency environment, investigating whether temporal aggregation can improve the bias bound by balancing signal and noise. In contrast, our work addresses the mixed-frequency data problem---situations in which donor series are observed at high, same, and low frequencies simultaneously---and proposes the MF-SCM. Rather than aggregating high-frequency outcomes, the MF-SCM integrates information across different temporal resolutions within a unified estimation framework. Consequently, methods based on disaggregated outcomes in \citet{sun2024temporal} cannot be directly applied in our setting, as the treated unit is observed at a lower frequency and hence lacks the high-frequency resolution required by their approach. The proposed MF-SCM generalizes their framework by enabling the joint integration of low-, same-, and high-frequency data. Moreover, whereas \citet{sun2024temporal} derive finite-sample bias bounds, our results establish large-sample risk optimality of the estimated weights in a mixed-frequency environment.

\citet{zhu2023synthetic} derives a result similar to our Theorem \ref{th1} for a different estimator within the SCM framework.
While similar forms of asymptotic optimality results have appeared in the literature, our contribution lies in extending the synthetic control framework to accommodate mixed-frequency data and establishing that asymptotic optimality can also be achieved in this setting. Moreover, unlike \citet{zhu2023synthetic}, who proves in-sample risk optimality for the SRC estimator, Theorem \ref{th1} in our paper establishes the asymptotic optimality of the out-of-sample prediction risk, which is directly relevant to counterfactual estimation and the evaluation of post-treatment effects.

The remainder of the paper is organized as follows. Section \ref{sec:2} introduces the MF-SCM estimator and details its implementation. Section \ref{sec:3} establishes asymptotic properties of the MF-SCM estimator and discusses the assumptions required for asymptotic properties. Section \ref{sec:4} develops the distribution theory of the MF-SCM estimator, while Section \ref{sec:5} outlines its inference methodology.
Section \ref{sec:6} presents Monte Carlo simulation results. Section \ref{sec:7} evaluates the empirical performance of the MF-SCM by examining the effects of 2017 Tax Cuts and jobs Act in US and air pollution alerts in Beijing, and Section \ref{sec:8} draws some conclusions. Technical proofs are provided in the Appendix in the supplementary materials.

\section{SCM of mixed frequency data}\label{sec:2}
Suppose that we observe $J + 1$ units, where the first unit ($j = 1$) is designated as the treated unit, while the subsequent units ($j \in \{2, \ldots, J + 1\}$) serve as the control units.
Consider a mixed-frequency data sampling process $\{Y_{j,t}, \mathbf{X}_{j,t}\}$, where $Y_{j,t}$ represents the outcome of interest for unit $j$ at time $t$, which is sampled at different frequencies, and $\mathbf{X}_{j, t}$ denotes the $Q$-dimensional vector of covariates.
The treated unit's outcome $Y_{1,t}$ is observed at $t = 1, \ldots, T$, where $t$ indexes the basic time unit, meaning that $Y_{1,t}$ is observed once between $t-1$ and $t$ (i.e., between the time range $(t-1,t]$).
We categorize the control units into three groups based on their sampling frequency relative to $Y_{1,t}$. Specifically, we assume that $Y_{t}^{[1]} = \{Y_{j,t} : j \in \mathcal{J}_1\}~(\mathcal{J}_1 = \{2, \ldots, J_1\})$ is sampled at the same frequency as $Y_{1,t}$ (e.g., quarterly), $Y_{t}^{[2]} = \{Y_{j,t} : j\in \mathcal{J}_2\}~(\mathcal{J}_2 = \{J_{1}+1, \ldots, J_2\})$ is sampled at the lower frequency than $Y_{1,t}$ (e.g., yearly), and $Y_{t}^{[3]} = \{Y_{j,t} : j\in \mathcal{J}_3\}~(\mathcal{J}_3 = \{J_{2}+1, \ldots, J+1\})$ is sampled at the higher frequency (e.g., monthly).
The observations span $T$ basic time periods, with $T_{0}~(T_{0} < T)$ denoting the last period before the treatment in unit $j = 1$. Define $\mathcal{T}_0 = \{ 1, \ldots, T_0 \}$ as the pre-treatment periods and $\mathcal{T}_1 = \{T_0+1, \ldots, T_0+T_1\}$ as the post-treatment periods.
The potential outcomes are denoted as $Y_{j, t}^{I}$ when unit $j$ is treated at time $t$ and $Y_{j, t}^{N}$ when no treatment is applied.
A standard assumption in this setting is that the treatment does not affect outcomes during the pre-treatment period, ensuring that $Y_{j, t}^{N} = Y_{j, t}^{I}$ for all units $j$ and all pre-treatment periods $t \in \mathcal{T}_0$.

For the higher-frequency outcome, we denote it as $Y_{j, t}^{(m_j)}$ for $j\in \mathcal{J}_3$, where the superscript $m_j$ indicates that $Y_{j, t}^{(m_j)}$ is observed $m_j$ times within the interval $(t-1,t]$. Specifically, we represent these observations as a sequence $\{Y_{j, t-k/m_j}^{(m_j)} : k=0, \ldots, K_j\}$, where $K_j$ is the maximum lag order for the $j$-th unit. Throughout the paper, we take the sampling frequency of $Y_{1,t}$ as the baseline frequency. For notational simplicity, we assume that all covariates share a common sampling frequency across units and are observed at this baseline frequency.

To apply SCM to these mixed-frequency data, we develop a flexible estimation procedure to address the challenges posed by mixed-frequency data.

Specifically, for $j\in \mathcal{J}_2$, one straightforward approach is to treat the unobserved baseline-frequency outcomes as missing observations and model the relationship between \(Y_{j,t}\) and \(\mathbf{X}_{j,t}\) using a distributed lag (DL) models, which are widely used in the literature to capture the distribution over time of the lagged effects of changes in explanatory variable, to forecast missing $Y_{j,t}$ values effectively (e.g., \citeauthor{amemiya1967comparative}, \citeyear{amemiya1967comparative}).
In general, a stylized distributed lag model is defined as
$$Y_{j,t} = \alpha_0 + \sum_{p=0}^{P} \boldsymbol{\beta}_{j,p}^{\prime} \mathbf{X}_{j, t-p} + \epsilon_{j,t},$$
where $\alpha_0$ and $\boldsymbol{\beta}_{j,p}$ ($Q\times 1$) are unknown parameters, $P$ is fixed representing the number of baseline-frequency lags used in the temporal aggregation of $\mathbf{X}_{j, t}$, and $\epsilon_{j,t}$ is the idiosyncratic error term. Once the parameters $\alpha_0$ and $\boldsymbol{\beta}_{j,p}$ are estimated, we can construct predictions for $Y_{j,t}$ when $t\neq n \widetilde{m}_j$ ($n \in \mathbb{N}_{+}$), where $\widetilde{m}_j$ indicates that $Y_{j, t}~(j \in \mathcal{J}_2)$ was observed one time in $\widetilde{m}_j$ basic time units.

Such a treatment is appropriate for cases where the low-frequency outcome represents a sample taken at a specific point in time or where the data naturally align with this interpretation.
However, this assumption is restrictive and may not hold in more general situations. In many practical settings, the low-frequency outcome represents an aggregated indicator over the entire period rather than a single subperiod’s value. 
In such cases, directly treating, for example, an annual observation as equivalent to the last quarterly observation would be conceptually inconsistent and may lead to biased estimates.
To accommodate these more general cases, we adopt a latent-variable framework, in which the unobserved baseline-frequency outcomes $Y_{j,t}$ are treated as latent variables. Specifically, We posit that the (unobserved) baseline-frequency outcome satisfies
\begin{equation}
Y_{j,t} = f(\mathbf{X}_{j,t}) + \epsilon_{j,t},
\label{eq:latent_general}
\end{equation}
where $f(\cdot)$ captures the relationship between the outcome and covariates at the baseline frequency.
For analytical tractability, we adopt the distributed lag specification for $f(\cdot)$, namely,
\begin{equation}
Y_{j,t} = \alpha_0 + \sum_{p=0}^{P} \boldsymbol{\beta}_{j,p}^{\prime} \mathbf{X}_{j, t-p} + \epsilon_{j,t}.
\label{eq:latent_model}
\end{equation}
The observed lower-frequency outcome \(Y_{j, t^{(1:\widetilde{m}_j)}}\) is modeled as a weighted aggregation of these latent baseline-frequency outcomes:
\begin{equation}
Y_{j, t^{(1:\widetilde{m}_j)}} = \sum_{s=1}^{\widetilde{m}_j} W_{j,s} Y_{j, t-\widetilde{m}_j + s}, \quad t \in \{\widetilde m_j, 2\widetilde m_j, \ldots, \lfloor T/\widetilde m_j \rfloor \widetilde m_j\},
\label{eq:aggregation}
\end{equation}
where $\sum_{s=1}^{\widetilde{m}_j} W_{j,s} = 1$ to ensure identification. 
Each observed low-frequency outcome $Y_{j, t^{(1:\widetilde{m}_j)}}$ corresponds to a non-overlapping block of $\widetilde m_j$ consecutive baseline-frequency periods. 
Substituting (\ref{eq:latent_model}) into (\ref{eq:aggregation}) yields
\begin{equation}
Y_{j, t^{(1:\widetilde{m}_j)}} 
= \alpha_0 + \sum_{p=0}^{P} \boldsymbol{\beta}_{j,p}^\prime \sum_{s=1}^{\widetilde{m}_j} W_{j,s} \mathbf{X}_{j,t-\widetilde{m}_j + s - p} 
+ \sum_{s=1}^{\widetilde{m}_j} W_{j,s} \epsilon_{j,t-\widetilde{m}_j + s}.
\label{eq:combined_model}
\end{equation}
The parameters $(\alpha_0, \boldsymbol{\beta}_{j,p}, \mathbf{W}_j)$ can be jointly estimated based on (\ref{eq:combined_model}). Once these parameters are obtained, the latent baseline-frequency outcomes $Y_{j,t}$ can be reconstructed from the estimated relationship between $Y_{j,t}$ and $\mathbf{X}_{j,t}$. 
This representation facilitates efficient estimation of both the model parameters and the latent series, enabling the recovery of baseline-frequency outcomes $Y_{j,t}$ from observed covariates sampled at the same frequency.

For $j\in \mathcal{J}_3$, we adopt the MIDAS-style weighting approach to align the higher-frequency data with the frequency of $Y_{1,t}$. Specifically, we construct a baseline-frequency representation $\bar{Y}_{j,t}$, aligned with the frequency of $Y_{1,t}$, from the higher-frequency $Y_{j,t}^{(m_j)}$ via: 
$$\bar{Y}_{j,t} = \widetilde{B}_j(L^{1/m_j}; \boldsymbol{\zeta}_{j}) Y_{j,t}^{(m_j)} = \sum_{k=1}^{K_j} B_j(k; \boldsymbol{\zeta}_{j}) Y_{j,t-(k-1)/m_j}^{(m_j)},$$
where $\bar{Y}_{j,t}$ denotes a weighted aggregation of the higher-frequency outcome, $\widetilde{B}_j(L^{1/m_j}; \boldsymbol{\zeta}_{j}) = \sum_{k=1}^{K_j} B_j(k; \boldsymbol{\zeta}_{j})L^{(k-1)/m_j}$, and $L^{1/m_j}$ is the lag operator such that $L^{1/m_j}Y_{j,t}^{(m_j)} = Y_{j,t-1/m_j}^{(m_j)}$.
Here, we set $K_j = m_j$.

The parameterization of the lagged coefficients of $B_j(k; \boldsymbol{\zeta}_{j})$ in a parsimonious way is one of the key MIDAS features. One of the most used parameterizations is the ``exponential Almon lag'', which is closely related to the smooth polynomial Almon lag functions commonly applied in the distributed lag literature to reduce multicollinearity (\citeauthor{almon1965distributed}, \citeyear{almon1965distributed}). The exponential Almon lag is defined as:
\begin{equation}
B_j(k; \boldsymbol{\zeta}_{j}) = \frac{\exp \left(\zeta_{j1} k + \ldots + \zeta_{jl} k^l \right)} {\sum_{k=1}^{m_j} \exp \left(\zeta_{j1} k + \ldots + \zeta_{jl} k^l \right)}. \label{2.1}
\end{equation}
This function is known for its flexibility and can take various shapes with only a few parameters (e.g., \citeauthor{griffiths1985theory}, \citeyear{griffiths1985theory}, for further discussion), including decreasing, increasing or hump-shaped patterns. \citet{ghysels2005there} use the functional form (\ref{2.1}) with two parameters, which allows a great flexibility and determines how many lags are included in the regression.
Another parameterization, with only two parameters, is the so-called ``Beta Lag'', because it is based on the Beta function:
\begin{equation}
B_j(k;\boldsymbol{\zeta_j}) = \frac{f\left(\frac{k}{m_j}, \zeta_{j1}, \zeta_{j2}\right)}{\sum_{k=1}^{m_j} f\left(\frac{k}{m_j}, \zeta_{j1}, \zeta_{j1}\right)}, \notag
\end{equation}
where
$$f(x, a, b) = \frac{x^{a-1}(1-x)^{b-1} \Gamma(a+b)}{\Gamma(a) \Gamma(b)},$$
and
$$\Gamma(a) = \int_0^{\infty} e^{-x} x^{a-1} dx.$$
The parameterizations described above are all quite flexible. For different values of the parameters, they can take various shapes: weights attached to the different lags can decline slowly or fast, or even have a hump shape. 

To enable estimation while preserving convexity, we represent the MIDAS weight function in a fixed dictionary (basis), following the idea of \citet{babii2022machine}. Specifically, for a window length \(m_j\) and basis functions \(\{w^{(dic)}_\ell(\cdot)\}_{\ell=1}^L\) (e.g., Legendre, discrete splines, and orthogonal polynomials), we set
\begin{equation}
B_j(k;\boldsymbol{\zeta}_j)
=
\sum_{\ell=1}^{L_y} \zeta_{j,\ell}\, w^{(dic)}_\ell\!\Big(\tfrac{k-1}{m_j}\Big),
\qquad
k=1,\ldots,m_j.
\label{eq:dict-midas}
\end{equation}
This dictionary-based specification departs from the conventional parameterizations such as exponential Almon or Beta lags, yet it retains flexibility through the choice of basis functions while ensuring convexity in the optimization over \(\boldsymbol{\zeta}_j\). 
As noted by \citet{marsilli2014variable} and \citet{babii2022machine}, this representation leads to a computationally attractive convex optimization problem.

The interest is to estimate the treatment effect, defined as $\alpha_{j, t} = Y_{j, t}^{I} - Y_{j, t}^{N}$, for unit $j$ at time $t$. This  allows the observable outcome to be expressed using counterfactual notation as $Y_{j, t} = Y_{j, t}^{N} + \alpha_{j, t} D_{j, t}$, where $D_{j, t} = 1$ if $j=1$ and $t\in \mathcal{T}_1$, and $D_{j, t} = 0$ otherwise.
For the treated unit, the treatment effect in the post-treatment period is given by 
$$\alpha_{1,t} = Y_{1, t}^{I} - Y_{1, t}^{N} = Y_{1, t} - Y_{1, t}^{N}.$$
Thus, the crucial quantity to estimate is $Y_{1, t}^N$ for $t \in \mathcal{T}_1$, which represents the counterfactual outcome of the treated unit had it not received the treatment in the post-treatment periods.
To estimate $Y_{1, t}^N$, we first define $\widetilde{Y}_{j,t}^{N}$ as
$$
\widetilde{Y}_{j,t}^{N} = \begin{cases} Y_{j,t}, & \text { if } 2 \leq j \leq J_1, \\ \widehat{Y}_{j,t}, & \text { if } J_1 < j \leq J_2, \\ \sum_{k=1}^{m_j} B_j(k; \boldsymbol{\zeta}_j) Y_{j,t-(k-1)/m_j}^{(m_j)}, & \text { if } J_2 < j \leq J+1,\end{cases}
$$
where $\widehat{Y}_{j,t} = \widehat{\alpha}_0 + \sum_{p=0}^{P} \widehat{\boldsymbol{\beta}}_{j,p}^{\prime} \mathbf{X}_{j, t-p}$. 
Using these definitions, the MF-SCM constructs the estimation of $Y_{1, t}^{N}$ through an optimally weighting the outcomes of control units after aligning their mixed-frequency data, that is,
$$ \widehat{Y}_{1, t}^{N}(\mathbf{w}) = \sum_{j=2}^{J+1} w_j \widetilde{Y}_{j,t}^{N}, \quad t \in \mathcal{T}_1, $$
where $\mathbf{w} = (w_2, \ldots, w_{J+1})^{\prime}$ is the weight vector belonging to the set
$$\mathcal{H} = \left\{\mathbf{w}=\left(w_2, \ldots, w_{J+1}\right)^{\prime} \in[0,1]^J \mid \sum_{j=2}^{J+1} w_j = 1\right\}.$$

To incorporate covariate information, we define 
$$\bar{\mathbf{X}}_{j} = (\bar{\mathbf{X}}_{j,1}^{\prime}, \ldots, \bar{\mathbf{X}}_{j,Q}^{\prime})^{\prime},$$
where 
$$\bar{\mathbf{X}}_{j,q} = T_0^{-1} ( X_{j,q,1}, \ldots, X_{j,q,T_0} )^{\prime}.$$
For $j=1,\ldots,J_1$, we define
$$\mathbf{Z}_j = (Y_{j, 1}, \ldots, Y_{j, T_{0}}, \bar{\mathbf{X}}_{j}^{\prime} )^{\prime}.$$
For $j \in \mathcal{J}_2$, we construct $\mathbf{Z}_j$ as
$$\mathbf{Z}_j = (\widehat{Y}_{j, 1}, \ldots, \widehat{Y}_{j, T_{0}}, \bar{\mathbf{X}}_{j}^{\prime} )^{\prime}.$$
For $j\in \mathcal{J}_3$, let
$$\mathbf{Z}_j = ( \sum_{k=1}^{m_j} B_j(k; \boldsymbol{\zeta}_j) Y_{j,1-(k-1)/m_j}^{(m_j)}, \ldots, \sum_{k=1}^{m_j} B_j(k; \boldsymbol{\zeta}_j) Y_{j,T_{0} - (k-1)/m_j}^{(m_j)}, \bar{\mathbf{X}}_{j}^{\prime} )^{\prime}.$$  
Finally, we define $\mathbb{Z}(\boldsymbol{\theta}) = (\mathbf{Z}_2, \ldots, \mathbf{Z}_{J+1})$, with $\boldsymbol{\theta} = (\boldsymbol{\zeta}_{J_2+1}^\prime,\ldots,\boldsymbol{\zeta}_{J+1}^\prime)^\prime$.

For any positive definite matrix $\mathbb{V}$, the weight of MF-SCM can be obtained by solving the following optimization:
$$(\widehat{\mathbf{w}}(\mathbb{V}), \widehat{\boldsymbol{\theta}}) = \arg \min _{\mathbf{w} \in \mathcal{H}, \boldsymbol{\theta} \in \mathcal{R}^{M}} \left(\mathbf{Z}_1 - \mathbb{Z}(\boldsymbol{\theta}) \mathbf{w}\right)^{\prime} \mathbb{V} \left(\mathbf{Z}_1 - \mathbb{Z}(\boldsymbol{\theta}) \mathbf{w}\right),$$
where $M=(J+1-J_2)l_1$, with $l_1$ depending on the parameterization chosen for $B_j(k; \boldsymbol{\zeta}_{j})$. For the sake of simplicity, we follow \citet{ferman2021synthetic} and set $\mathbb{V} = \mathbb{I}_{T_z}$, where $\mathbb{I}_{T_z}$ denotes the $T_z \times T_z$ identity matrix and $T_z \equiv (Q + 1) T_0$. To simplify notation, we write $\widehat{\mathbf{w}}(\mathbb{I}_{T_z})$ as $\widehat{\mathbf{w}}$. Then, the MF-SCM weight can be obtained by
\begin{eqnarray}
(\widehat{\mathbf{w}}, \widehat{\boldsymbol{\theta}}) & = & \arg \min _{\mathbf{w} \in \mathcal{H}, \boldsymbol{\theta}\in \mathcal{R}^{M}} L_{T_0}(\mathbf{w}, \boldsymbol{\theta}) \nonumber\\
& = & \arg \min _{\mathbf{w} \in \mathcal{H}, \boldsymbol{\theta} \in \mathcal{R}^{M}} \frac{1}{T_0} \left\| \mathbf{Z}_1 - \mathbb{Z}(\boldsymbol{\theta}) \mathbf{w} \right\|^{2} \nonumber\\
& = & \arg \min _{\mathbf{w} \in \mathcal{H}, \boldsymbol{\theta} \in \mathcal{R}^{M}} \frac{1}{T_0} \left\{ \sum_{t=1}^{T_0} [ Y_{1, t} - ( \sum_{j=2}^{J_1} w_j Y_{j, t} + \sum_{j=J_1+1}^{J_2} w_j \widehat{Y}_{j, t} \right. \nonumber\\
&& + \left.\sum_{j=J_2+1}^{J+1} w_j \sum_{k=1}^{m_j} B_j(k; \boldsymbol{\zeta}_j) Y_{j,t-(k-1)/m_j}^{(m_j)} )]^2 + \|\bar{\mathbf{X}}_1 - \sum_{j=2}^{J+1} w_j \bar{\mathbf{X}}_j \|^{2} \right\}, \label{2.2}
\end{eqnarray}
where $L_{T_0}(\mathbf{w}, \boldsymbol{\theta}) = T_{0}^{-1} \left\| \mathbf{Z}_1 - \mathbb{Z}(\boldsymbol{\theta}) \mathbf{w} \right\|^{2}$ is the loss function, and $\| \cdot \|$  denotes the Euclidean norm. The corresponding risk function is defined as $R_{T_0}(\mathbf{w}, \boldsymbol{\theta}) = \mathbb{E} L_{T_0}(\mathbf{w}, \boldsymbol{\theta})$. 
We identify $\mathbf{w}$ and $\boldsymbol{\theta}$ by normalizing the $\mathbf{w}$ to sum to unity, and similarly normalize $B_j(L^{1/m_j}; \boldsymbol{\zeta}_j)$ so that its coefficients also sum to unity.

Then, the MF-SCM estimator of treatment effect is
$$\widehat{\alpha}_{1, t} (\widehat{\mathbf{w}}, \widehat{\boldsymbol{\theta}}) = Y_{1, t}^{I} - \sum_{j=2}^{J+1} \widehat{w}_j \widetilde{Y}_{j, t}^{N}, \quad t\geq T_0+1.$$

The optimization problem in (\ref{2.2}) involves two sets of parameters: the MIDAS parameters \(\boldsymbol{\theta}\) and the SCM weights \(\mathbf{w}\).
Under the dictionary-based specification, the optimization with respect to the MIDAS parameters \(\boldsymbol{\theta}\) is convex. Moreover, for fixed MIDAS parameters, the optimization with respect to the SCM weights \(\mathbf{w}\) constitutes a convex quadratic program, as the loss function \(L_{T_0}(\mathbf{w}, \boldsymbol{\theta})\) is quadratic in \(\mathbf{w}\) and the feasible set \(\mathcal{H}\) is convex. 
However, the joint optimization problem in (\ref{2.2}) over \((\mathbf{w}, \boldsymbol{\theta})\) is not necessarily convex, since the two sets of parameters are coupled within the objective function. 
To ensure that the proposed estimator can still be computed via convex optimization, we further examine the convexity properties of the joint optimization problem in (\ref{2.2}) and show that it can be equivalently reformulated as a convex optimization problem. 
A detailed discussion of these convexity properties is provided in Appendix F.

Next, we examine the role of the covariate $\mathbf{X}_{j,t}$ in weight optimization. First, we rewrite (\ref{2.2}) as
\begin{eqnarray}
(\widehat{\mathbf{w}}, \widehat{\boldsymbol{\theta}}) & = & \arg \min _{\mathbf{w} \in \mathcal{H}, \boldsymbol{\theta} \in \mathcal{R}^{M}} \frac{1}{T_0} \left\{ \sum_{t=1}^{T_0} ( Y_{1, t} - \sum_{j=2}^{J+1} w_j \widetilde{Y}_{j, t}^{N} )^{2} + \|\bar{\mathbf{X}}_1 - \sum_{j=2}^{J+1} w_j \bar{\mathbf{X}}_j \|^{2}\right\}  \nonumber\\
& = & \arg \min _{\mathbf{w} \in \mathcal{H}, \boldsymbol{\theta} \in \mathcal{R}^{M}} \frac{1}{T_0} \left\{ \sum_{t=1}^{T_0} ( Y_{1, t} - \sum_{j=2}^{J+1} w_j \widetilde{Y}_{j, t}^{N} )^{2}  \right. \nonumber\\
&& +\left. \sum_{q=1}^{Q} \sum_{t=1}^{T_0} \frac{1}{T_0} \left( X_{1, q, t}  - \sum_{j=2}^{J+1} w_j X_{j,q,t} \right)^2 \right\}. \label{2.3}
\end{eqnarray}
In the weight estimation procedure specified by (\ref{2.3}), the normalization of each component in the $\bar{\mathbf{X}}_j~(j = 1, \ldots, J+1)$ by $T_0$ ensures that the relative importance of covariate information in weight optimization remains consistent with the classical SCM framework. This scaling operation preserves the invariant contribution ratio of covariates to the optimization objective. 
Notably, when the covariate $\mathbf{X}_{j,t}$ is time-invariant characteristics, the term $\|\bar{\mathbf{X}}_1 - \sum_{j=2}^{J+1} w_j \bar{\mathbf{X}}_j \|^{2}$ in (\ref{2.3}) degenerates to its counterpart in the classical SCM. 

In summary, the algorithm for MF-SCM is shown in Algorithm \ref{alg:mfscm}.

\begin{algorithm}[htbp]
\caption{MF-SCM}
\label{alg:mfscm}
\begin{algorithmic}[1]
\STATE \textbf{Input:} Pre-treatment sample $\{Y_{j,t}, \mathbf{X}_{j,t}\}$, frequency partition $(\mathcal{J}_1,\mathcal{J}_2,\mathcal{J}_3)$, and post-treatment treated outcomes.
\FOR{each $j\in\mathcal{J}_2$}
\STATE estimate the latent reconstruction parameters from (\ref{eq:combined_model}) and recover the baseline-frequency latent series $\widehat{Y}_{j,t}$.
\ENDFOR
\FOR{each $j\in\mathcal{J}_3$}
\STATE construct aligned outcomes using the dictionary parameterization in (\ref{eq:dict-midas}).
\ENDFOR
\STATE Build $\widetilde{Y}_{j,t}^{N}$ and $\bar{\mathbf X}_j$ for all control units, then form the pre-treatment objective in (\ref{2.3}).
\STATE Solve the convex reformulation of (\ref{2.3}) (Appendix F) to obtain $(\widehat{\mathbf w},\widehat{\boldsymbol\theta})$.
\FOR{$t\in\mathcal{T}_1$}
\STATE Compute the counterfactual $\widehat{Y}_{1,t}^N=\sum_{j=2}^{J+1}\widehat{w}_j\widetilde{Y}_{j,t}^{N}$ and treatment effects $\widehat{\alpha}_{1,t}=Y_{1,t}^{I}-\widehat{Y}_{1,t}^{N}$.
\ENDFOR
\STATE \textbf{Output:} $(\widehat{\mathbf w},\widehat{\boldsymbol\theta})$, $\widehat{Y}_{1,t}^{N}$, and $\widehat{\alpha}_{1,t}$.
\end{algorithmic}
\end{algorithm}

\section{Asymptotic Optimality}\label{sec:3}
In this section, we will list some assumptions and present our main theoretical result. Unless otherwise stated, all limiting properties throughout the text are with respect to $T_0 \rightarrow \infty$.

To evaluate the performance of the MF-SCM estimator of $Y_{1,t}^{N}$, we consider the mean squared prediction error defined as
\begin{eqnarray*}
L_{T_1}(\mathbf{w}, \boldsymbol{\theta}) & = & \frac{1}{T_1} \sum_{t \in \mathcal{T}_1} (Y_{1, t}^{N} - \sum_{j=2}^{J+1} w_j \widetilde{Y}_{j,t})^2.
\end{eqnarray*}

Denote $R_{T_1}(\mathbf{w}, \boldsymbol{\theta}) = \mathbb{E} L_{T_1}(\mathbf{w}, \boldsymbol{\theta})$, $\xi_{T_0} = \inf _{\mathbf{w} \in \mathcal{H}, \boldsymbol{\theta}} R_{T_0}(\mathbf{w}, \boldsymbol{\theta})$, and $\xi_{T_1} = \inf _{\mathbf{w} \in \mathcal{H}, \boldsymbol{\theta}} R_{T_1}(\mathbf{w}, \boldsymbol{\theta})$.
To show the asymptotic optimality, we state some assumptions. All explanations of these assumptions are given below. 

\begin{assumption} \label{ass1}
{\it $ \xi_{T_0}^{-1} {T_0}^{-1/2} J^{2} = o(1)$.}
\end{assumption}

\begin{assumption} \label{ass2}
{\it There exists a constant $C_0$ such that $\mathbb{E} \|\mathbf{X}_j\| < C_0$ for $j \in \{1,\ldots, J+1\}$.}
\end{assumption}

Assumption \ref{ass1} imposes restrictions on the relative rate of several quantities approaching infinity, i.e., $\xi_{T_0},~T_0$ and $J$.
It is crucial to highlight that this assumption implies $\xi_{T_0} \neq 0$, a crucial assumption for establishing the asymptotic optimality of the MF-SCM weight. Intuitively, $\xi_{T_0} \neq 0$ means that it is impossible to achieve a perfect fit for the pre-treatment outcomes and observed covariates of the treated unit using a linear combination of the outcomes and covariates of the control units, and this situation is referred to as an imperfect pre-treatment fit. 
Assumption \ref{ass2} requires that the expectations of norms of covariates are uniformly bounded across all units, ensuring that $\mathbb{E} \|\mathbf{X}_j\|$ does not diverge.

\begin{assumption} \label{ass3}
{\it $\sup _{\mathbf{w} \in \mathcal{H}, \boldsymbol{\theta}} \left| T_0^{-1} \sum_{t \in \mathcal{T}_0} \mathbb{E} ( Y_{1, t} - \sum_{j=2}^{J+1} w_j \widetilde{Y}_{j, t}^{N} )^{2} - R_{T_1} (\mathbf{w}, \boldsymbol{\theta}) \right| = O({T_0}^{-1/2} J^2) \allowbreak + o(\xi_{T_0})$.}
\end{assumption}

Assumption \ref{ass3} imposes constraints on the discrepancy between the pre-treatment and post-treatment fit, implying that the primary distinction in the outcomes between the pre-treatment and post-treatment periods is fully due to the treatment effect.
A similar assumption has been discussed in \citet{hansen2012jackknife}.
To better illustrate Assumption \ref{ass3}, we use the distributed lag model as a special case. For units observed at the baseline frequency, suppose that the untreated potential outcomes $Y_{j,t}^N$ satisfy
\begin{eqnarray}
Y_{j,t}^N = \alpha_0 + \sum_{p=0}^{P} \boldsymbol{\beta}_{j,p}^{\prime} \mathbf{X}_{j, t-p} + \epsilon_{j,t}. \label{4.1}
\end{eqnarray}
Since Assumption \ref{ass3} involves the aligned outcomes $\widetilde{Y}_{j,t}^N$, it is also necessary to specify the data-generating process for units observed at a higher frequency. 
Specifically, for $j \in \mathcal{J}_3$, suppose that the untreated potential outcomes satisfy
\begin{eqnarray}
Y_{j,t-(k-1)/m_j}^{(m_j)} = \alpha^{(h)}_0 + \sum_{p=0}^{P} \boldsymbol{\beta}_{j,p}^{(h) \prime} \mathbf{X}_{j, t-(k-1)/m_j-p/m_j} + \epsilon^{(h)}_{j,t-(k-1)/m_j}. \label{4.1h}
\end{eqnarray}
Then, Assumption \ref{ass3} can be derived from more general assumptions as follows. The detailed proof is provided in Appendix D.

\begin{subassumption} \label{ass3.1}
{\it (i) We treat $\{\alpha_0, \alpha^{(h)}_0, \boldsymbol{\beta}_{j,p}, \boldsymbol{\beta}^{(h)}_{j,p} ~|~p\in \{0,1, \ldots, P \} \}$ as fixed, and $\{ \mathbf{X}_{j,t-p}, \mathbf{X}_{j,t-(k-1+p)/m_j}~|~j\in \{1,\ldots, J+1 \}, ~t \in \mathcal{T}_0 \cup \mathcal{T}_1,~p \in \{ 0,1,\ldots,P\},~k\in \{1, \ldots, m_j \} \}$ and $\{\epsilon_{j,t}~|~j\in \{1,\ldots, J+1 \}, ~t \in \mathcal{T}_0 \cup \mathcal{T}_1 \}$ as stochastic.}

{\it (ii) $\mathbb{E}\epsilon_{j,t} = 0$ and $\mathbb{E}\epsilon^{(h)}_{j,t-(k-1)/m_j} = 0$ for $j \in \{1,\ldots, J+1 \}$, $k\in \{1, \ldots, m_j \}$ and $t \in \mathcal{T}_0 \cup \mathcal{T}_1$.}
\end{subassumption}

\begin{subassumption}\label{ass3.2}
{\it For any $p \in \{0,1, \ldots, P \}$ and $q \in \{1, \ldots, Q \}$, there exists a constant $C_1$ such that: 
(i) $\alpha_0 < C_1$, $\alpha^{(h)}_0 < C_1$, $\beta_{j, p, q} < C_1$ and $\beta^{(h)}_{j, p, q} < C_1$; 
(ii) $\widehat{\alpha}_0 < C_1$, $\widehat{\alpha}^{(h)}_0 < C_1$, $\widehat{\beta}_{j, p,q} < C_1$ and $\widehat{\beta}^{(h)}_{j, p,q} < C_1$ almost surely.}
\end{subassumption}

\begin{subassumption}\label{ass3.3}
{\it $T_0^{-1} \sum_{t \in \mathcal{T}_{0}} \mathbb{E}(\mathbf{X}_{i, t - (k_1-1)/m_i}^{\prime} \mathbf{X}_{j, t - (k_2-1)/m_j}) - T_1^{-1} \sum_{t \in \mathcal{T}_{1}} \mathbb{E}(\mathbf{X}_{i, t - (k_1-1)/m_i}^{\prime} \mathbf{X}_{j, t - (k_2-1)/m_j}) = O( T_0^{-1/2} )$ for $i \in \{1,\ldots, J+1 \}$, $j \in \{1,\ldots, J+1 \}$ and $k_1, k_2 \in \{1, \ldots, K_0 \}$.}
\end{subassumption}

We also need some restrictions on the relation between the idiosyncratic shock of the treated and control units. Let $\tilde{e}_{t, \epsilon}^{(j)} = \epsilon_{1, t} - \widetilde{\epsilon}_{j, t}$ for $j = 2, \ldots, J+1 $ and $t \in \mathcal{T}_0 \cup \mathcal{T}_1$, where
$$
\widetilde{\epsilon}_{j, t} = \begin{cases} \epsilon_{j, t}, & \text { if } 2 \leq j \leq J_1, \\ 0, & \text { if } J_1 < j \leq J_2, \\ \sum_{k=1}^{K_0} B_j(k; \boldsymbol{\zeta}_j) \epsilon^{(h)}_{j,t-(k-1)/m_j}, & \text { if } J_2 < j \leq J+1. \end{cases}
$$

\begin{subassumption}\label{ass3.4}
{\it $\sup _{\mathbf{w} \in \mathcal{H}, \boldsymbol{\theta}_0}\left| T_0^{-1} \sum_{t \in \mathcal{T}_0} \mathbb{E} \left(\sum_{j=1}^J w_j \tilde{e}_{t, \epsilon}^{(j)} \right)^2 - T_1^{-1} \sum_{t \in \mathcal{T}_1} \mathbb{E} \left(\sum_{j=1}^J w_j \tilde{e}_{t, \epsilon}^{(j)}\right)^2 \right| = o(\xi_{T_0})$.}
\end{subassumption}

Assumption \ref{ass3.1} (i) is introduced to simplify the proof, and a similar assumption is also employed in \citet{ferman2021properties} and \citet{ferman2021synthetic}. Assumption \ref{ass3.2} (ii) imposes a zero-mean condition on the idiosyncratic shocks, a standard assumption in the SCM literature (e.g., \citeauthor{botosaru2019role}, \citeyear{botosaru2019role}; \citeauthor{ferman2021properties}, \citeyear{ferman2021properties}, among others).
Assumption \ref{ass3.2} requires the uniform boundedness of the parameters $\alpha_0$ and $\boldsymbol{\beta}_{j,p}$ and the uniform boundedness of their estimators of $\widehat{\alpha}_0$ and $\widehat{\boldsymbol{\beta}}_{j,p}$, respectively.
Assumption \ref{ass3.3} implies that while covariates may differ between the pre-treatment and post-treatment periods, this discrepancy gradually diminishes as the number of pre-treatment periods increase. This suggests that the variation in covariates does not deviate significantly after treatment, indicating that the treatment effect alone accounts for the majority of the difference in the outcomes between the pre-treatment and post-treatment periods.
Assumption \ref{ass3.4} means that, with respect to the goodness of pretreatment fit, the difference in idiosyncratic shocks between the treated and any weighted average of control units does not change substantially after treatment.

To further illustrate Assumption \ref{ass3}, we consider the following factor model for units observed at the baseline frequency, as used in \citet{ferman2021properties}, 
\begin{align}
Y_{j,t}^N = \boldsymbol{\lambda}_{j}^{\prime} \boldsymbol{f}_{t} + \varepsilon_{j,t}, \label{m.1}
\end{align}
where $\boldsymbol{f}_t = (f_{1,t}, \ldots, f_{F,t})^{\prime}$ is an $F \times 1$ vector of unobserved common factors, $\boldsymbol{\lambda}_{j}$ is an $F \times 1$ vector of unknown factor loadings, and $\varepsilon_{j,t}$ is an idiosyncratic shock.
It is also necessary to specify the data-generating process for units observed at a higher frequency. Specifically, for $j \in \mathcal{J}_3$, suppose that the untreated high-frequency potential outcomes satisfy
\begin{align}
Y_{j,t-(k-1)/m_j}^{(m_j)} = \boldsymbol{\lambda}_{j}^{(h) \prime} \boldsymbol{f}_{t-(k-1)/m_j} + \varepsilon^{(h)}_{j,t-(k-1)/m_j},
\label{m.1h}
\end{align}

We now demonstrate that Assumption \ref{ass3} can be derived from more general assumptions as follows. The detailed proof showing how these sub-assumptions jointly imply Assumption \ref{ass3} is provided in Appendix E.

\begin{assumptionprime} \label{ass3.1 prime}
{\it (i) We treat $\{\boldsymbol{\lambda}_{j} ~|~j\in \{1, \ldots, J+1\} \}$ and $\{\boldsymbol{f}_{t} ~|~t\in \mathcal{T}_0 \cup \mathcal{T}_1\}$ as fixed, and $\{\varepsilon_{j,t}~|~j\in \{1,\ldots, J+1 \}, ~t \in \mathcal{T}_0 \cup \mathcal{T}_1 \}$ as stochastic.}

{\it (ii) $\mathbb{E}\varepsilon_{j,t} = 0$ for $j \in \{1,\ldots, J+1 \}$ and $t \in \mathcal{T}_0 \cup \mathcal{T}_1$.}
\end{assumptionprime}

\begin{assumptionprime} \label{ass3.2 prime}
{\it There exists a constant $C_0$ such that $\|\boldsymbol{\lambda}_{j}\| < C_1$ and $\|\boldsymbol{f}_{t}\| < C_1$ for $j \in \{1, \ldots, J+1 \}$ and $t \in \mathcal{T}_0 \cup \mathcal{T}_1$.}
\end{assumptionprime}

\begin{assumptionprime} \label{ass3.3 prime}
{\it $T_0^{-1} \sum_{t \in \mathcal{T}_{0}} \boldsymbol{f}_{t - (k_1-1)/m_i}^{\prime} \boldsymbol{f}_{t - (k_2-1)/m_j} - T_1^{-1} \sum_{t \in \mathcal{T}_{1}} \boldsymbol{f}_{t - (k_1-1)/m_i}^{\prime} \boldsymbol{f}_{t - (k_2-1)/m_j} = O( T_0^{-1/2} )$ for $i \in \{1,\ldots, J+1 \}$, $j \in \{1,\ldots, J+1 \}$ and $k_1, k_2 \in \{1, \ldots, K_0 \}$.}
\end{assumptionprime}

\begin{assumptionprime} \label{ass3.4 prime}
{\it $\mathbb{E} \|T_0^{-1} \sum_{t_0 \in \mathcal{T}_{0}} f_{l,t_0 - (k_1-1)/m_i} \mathbf{X}_{j, t_0 - (k_2-1)/m_j} - T_1^{-1} \sum_{t_1 \in \mathcal{T}_{1}} f_{l,t_1 - (k_1-1)/m_i} \mathbf{X}_{j, t_1 - (k_2-1)/m_j} \| = O( T_0^{-1/2} )$ for $i,j \in \{1,\ldots, J+1 \}$, $l \in \{1,\ldots, F \}$ and $k_1, k_2 \in \{1, \ldots, K_0 \}$.}
\end{assumptionprime}

Let $\tilde{e}_{t, \varepsilon}^{(j)} = \varepsilon_{1, t} - \widetilde{\varepsilon}_{j, t}$ for $j = 2, \ldots, J+1 $ and $t \in \mathcal{T}_0 \cup \mathcal{T}_1$, where
$$
\widetilde{\varepsilon}_{j, t} = \begin{cases} \varepsilon_{j, t}, & \text { if } 2 \leq j \leq J_1, \\ 0, & \text { if } J_1 < j \leq J_2, \\ \sum_{k=1}^{K_0} B_j(k; \boldsymbol{\zeta}_j) \varepsilon^{(h)}_{j,t-(k-1)/m_j}, & \text { if } J_2 < j \leq J+1. \end{cases}
$$

\begin{assumptionprime} \label{ass3.5 prime}
{\it $\sup _{\mathbf{w} \in \mathcal{H}, \boldsymbol{\theta}_0}\left| T_0^{-1} \sum_{t \in \mathcal{T}_0} \mathbb{E} \left(\sum_{j=1}^J w_j \tilde{e}_{t, \varepsilon}^{(j)} \right)^2 - T_1^{-1} \sum_{t \in \mathcal{T}_1} \mathbb{E} \left(\sum_{j=1}^J w_j \tilde{e}_{t, \varepsilon}^{(j)}\right)^2 \right| = o(\xi_{T_0})$.}
\end{assumptionprime}

Analogous to Assumption \ref{ass3.1}, Assumption \ref{ass3.1 prime} (i) is introduced to simplify the proof, and Assumption \ref{ass3.1 prime} (ii) imposes a zero-mean condition on the idiosyncratic shock.
Assumption \ref{ass3.2 prime} requires that both factor loadings and common factors are uniformly bounded.
Assumption \ref{ass3.3 prime} requires that the variation in common factors does not change substantially after treatment, implying that the primary difference between the pre- and post-treatment outcomes is attributable solely to the treatment effect. Assumption \ref{ass3.4 prime} ensures that the cross-moment structure between factor components and the covariates remains stable across the pre-treatment period $\mathcal{T}_0$ and the post-treatment period $\mathcal{T}_1$. Assumption \ref{ass3.5 prime} serves the same purposes as Assumption \ref{ass3.4}.

\begin{assumption} \label{ass4}
{\it For any $i \in\{1, \ldots, J+1\}$ and $t \in \mathcal{T}_0 \cup \mathcal{T}_1$, $\{Y_{i,t}^{N}\}$ is either $\alpha$-mixing with the mixing coefficient $\alpha = -r /(r-2)$ or $\phi$-mixing with the mixing coefficient $\phi = -r /(2 r-1)$ for $r \geq 2$.}
\end{assumption}

Denote $e_{t, \widetilde{Y}^{N}}^{(i)} = Y_{1, t} - \widetilde{Y}_{i,t}$ for $i \in\{2, \ldots, J+1\}$ and $t \in \mathcal{T}_0 \cup \mathcal{T}_1$.

\begin{assumption} \label{ass5}
{\it (i) There exists a constant $C_2$ such that $\mathbb{E} (Y_{i,t}^{N})^4 \leq C_2 < \infty$ for $i \in\{1, \ldots, J+1\}$ and $t \in \mathcal{T}_0 \cup \mathcal{T}_1$.}

{\it (ii) There exists a constant $C_3$ such that $\operatorname{var} ( T_0^{-1/2} \sum_{t=1}^{T_0} e_{t, \widetilde{Y}^{N}}^{(i)} e_{t, \widetilde{Y}^{N}}^{(j)} ) \geq C_3 > 0$ for all $T_0$ sufficiently large and any $i, j \in\{2, \ldots, J+1\}$, and for any $\boldsymbol{\theta}$.}
\end{assumption}

Assumption \ref{ass4} imposes constraints on the dependency of the potential outcomes $Y_{i,t}^{N}$. Assumption \ref{ass5} (i) implies that the fourth moments of all $Y_{i, t}^{N}$ can be uniformly bounded. Assumption \ref{ass5} (ii) concerns the difference between the potential outcomes of the treated and control units, ensuring that these variances do not degenerate as $T_0$ increase.

\begin{theorem} \label{th1}
{\it Under Assumptions \ref{ass1}-\ref{ass5}, we have
\begin{eqnarray}
\frac{R_{T_1}(\widehat{\mathbf{w}}, \widehat{\boldsymbol{\theta}})}{\inf _{\mathbf{w} \in \mathcal{H}, \boldsymbol{\theta} \in \mathcal{R}^{M}} R_{T_1}(\mathbf{w}, \boldsymbol{\theta})} \stackrel{p}{\rightarrow} 1. \label{3.1}
\end{eqnarray}
}
\end{theorem}

Theorem \ref{th1} establishes the asymptotic optimality of the MF-SCM estimator. Specifically, (\ref{3.1}) shows that the MF-SCM weight is asymptotically optimal among all possible weighting combinations in the sense that the risk $R_{T_1}(\widehat{\mathbf{w}}, \widehat{\boldsymbol{\theta}})$ are asymptotically identical to those of the infeasible but best estimator.

\section{Distribution Theory}\label{sec:4}
This section explores the distribution theory associated with the proposed Mixed-Frequency SC method.
The asymptotic analysis of estimators under convex constraints is closely related to the literature on boundary asymptotics and constrained estimation. 
When the true parameter lies on the boundary of the parameter space, the limiting distribution of a constrained estimator is generally nonstandard and can often be characterized through projections onto tangent cones; see, for example, \citet{andrews1999estimation} and \citet{geyer1994asymptotics}. 
Projection-based arguments for convex sets are also closely related to \citet{zarantonello1971projections}.

We begin by considering the treatment effects $\alpha_{j,t}$, which are assumed to follow a weakly stationary process. Under this assumption, we can define the average treatment effects (ATE) for unit $j$ as $\alpha_j = \mathbb{E}(\alpha_{j,t})$, where
the expectation is taken with respect to the stationary distribution of $\alpha_{j,t}$. The ATE of treated unit can be estimated using the following expression:
$$\widehat{\alpha}_1 = T_1^{-1} \sum_{t=T_0+1}^T (Y_{1,t} - \sum_{j=2}^{J+1} \widehat{w}_j \widetilde{Y}_{j, t}^{N}).$$

After replacing $\boldsymbol{\theta}$ with $\widehat{\boldsymbol{\theta}}$ obtained in (\ref{2.3}),
the problem of finding the MF-SCM weight can be reformulated as a constrained regression problem based on the following regression model:
\begin{eqnarray}
Z_{1, \tilde{t}} = \widetilde{\mathbf{Z}}^{\prime}_{\tilde{t}} \mathbf{w}_0 + u_{\tilde{t}}, \label{5.1}
\end{eqnarray}
where $\widetilde{\mathbf{Z}}_{\tilde{t}} \equiv (\widetilde{Z}_{2, \tilde{t}}, \ldots, \widetilde{Z}_{J+1, \tilde{t}})^{\prime}$ ($\tilde{t} = 1, \ldots, T_z$) is a $J \times 1$ vector representing the $\tilde{t}$-th row of $\widetilde{\mathbb{Z}}$, and the matrix $\widetilde{\mathbb{Z}}$ is constructed by replacing $\boldsymbol{\theta}$ in $\mathbb{Z}(\boldsymbol{\theta})$ with $\widehat{\boldsymbol{\theta}}$, which is the $T_z \times J$-matrix with entry $\widetilde{Z}_{j, \tilde{t}}$ at position $(\tilde{t}, j-1)$, $\mathbf{w}_0 = (w_{0,2}, \ldots, w_{0,J+1})^{\prime}$ is an $J × 1$ vector of unknown coefficients, and we assume that the true parameter $\mathbf{w}_0$ belongs to $\mathcal{H}$, i.e., $\mathbf{w}_0 \in \mathcal{H}$,
and the term $u_{\tilde{t}}$ is a zero mean, finite variance idiosyncratic error term.
The weights $\mathbf{w} = (w_2, \ldots, w_{J+1})^{\prime}$ is then selected via the following constrained minimization problem:
\begin{eqnarray*}
\widehat{\mathbf{w}}_{\rm MF-SCM} = \arg \min _{\mathbf{w} \in \mathcal{H}} \frac{1}{T_0} \left\| \mathbf{Z}_1 - \widetilde{\mathbb{Z}} \mathbf{w} \right\|^{2},
\end{eqnarray*}
Notably, $\widehat{\mathbf{w}}_{\rm MF-SCM}$ is equivalent to $\widehat{\mathbf{w}}$ obtained from $(\ref{2.3})$.
\citeauthor{hsiao2012panel}'s (\citeyear{hsiao2012panel}) suggested estimating $\mathbf{w}$ in model (\ref{5.1}) with the least squares method. Let $\widehat{\mathbf{w}}_{\rm OLS}$ denote the ordinary least squares (OLS) estimator of $\mathbf{w}$ using the pre-treatment period data, that is,
\begin{eqnarray}
\widehat{\mathbf{w}}_{\rm OLS} = \arg \min _{\mathbf{w} \in \mathcal{R}^{J}} \frac{1}{T_0} \left\| \mathbf{Z}_1 - \widetilde{\mathbb{Z}} \mathbf{w} \right\|^{2}. \label{5.2}
\end{eqnarray}

To derive the distribution theory of the ATE estimator, we begin by demonstrating that the constrained MF-SCM estimator can be expressed as a projection of the unconstrained OLS estimator onto a constrained set. Using the theory of projection onto convex sets, we then derive the asymptotic distribution of the ATE estimator.

Let $\widetilde{\mathbb{Y}} = (\widetilde{\mathbf{Y}}_{1}, \ldots, \widetilde{\mathbf{Y}}_{T_0})^{\prime}$ be a $T_0\times J$ matrix formed by the first $T_0$ rows of $\widetilde{\mathbb{Z}}$, where $\widetilde{\mathbf{Y}}_{t}$ denotes its $t$-th row, i.e.,
\begin{align*}
\widetilde{\boldsymbol{Y}}_t = \bigg( &Y_{2, t}, \ldots, Y_{J_1, t}, \widehat{Y}_{J_1+1,t}, \ldots, \widehat{Y}_{J_2,t}, 
\sum_{k=1}^{m_{J_2+1}} B_{J_2+1}(k; \widehat{\boldsymbol{\zeta}}_{J_2+1}) Y_{J_2+1,t-(k-1)/m_{J_2+1}}^{(m_{J_2+1})}, \notag \\
&\ldots, \sum_{k=1}^{m_{J+1}} B_{J+1}(k; \widehat{\boldsymbol{\zeta}}_{J+1}) Y_{J+1,t-(k-1)/m_{J+1}}^{(m_{J+1})} \bigg)^{\prime},
\end{align*}
and let $\widetilde{\mathbb{X}}$ be a $(T_z-T_0) \times J$ matrix formed by the remaining $T_z - T_0 $ rows of $\widetilde{\mathbb{Z}}$.
For $\tau \in \mathcal{R}^J$, we define two versions of projection of $\tau$ onto a convex set $\mathcal{H}$ as follows:
$$\Pi_{\mathcal{H}, T_0} \tau = \arg \min _{\lambda \in \mathcal{H}} (\tau-\lambda)^{\prime} (\widetilde{\mathbb{Z}}^{\prime} \widetilde{\mathbb{Z}}/T_0) (\tau-\lambda),$$
$$\Pi_{\mathcal{H}} \tau = \arg \min _{\lambda \in \mathcal{H}} (\tau-\lambda)^{\prime} \mathbb{E}(\widetilde{\mathbf{Y}}_{t} \widetilde{\mathbf{Y}}_{t}^{\prime}) (\tau-\lambda).$$
Here, $\Pi_{\mathcal{H}}$ denotes a projection onto the set $\mathcal{H}$.
The first projection, $\Pi_{\mathcal{H}, T_0}$, is defined with respect to a random norm $\|a\|_{\widetilde{\mathbb{Z}}} = \sqrt{a^{\prime} (\widetilde{\mathbb{Z}}^{\prime} \widetilde{\mathbb{Z}}/T_0) a}$, where $\widetilde{\mathbb{Z}} $ is a data-dependent matrix. In contrast, the second projection, $\Pi_{\mathcal{H}}$, is defined with respect to a nonrandom norm $\|a\|_{\mathbb{E}} = \sqrt{a^{\prime} \mathbb{E}(\widetilde{\mathbf{Y}}_{t} \widetilde{\mathbf{Y}}_{t}^{\prime}) a}$, that is, $\Pi_{\mathcal{H}, T_0} \tau = \arg \min _{\lambda \in \mathcal{H}} \|\tau-\lambda\|_{\widetilde{\mathbb{Z}}}^2$ and $\Pi_{\mathcal{H}} \tau = \arg \min_{\lambda \in \mathcal{H}} \|\tau-\lambda\|_\mathbb{E}^2$.
These projections establish connections between $\widehat{\mathbf{w}}$ and $\widehat{\mathbf{w}}_{\mathrm{OLS}}$, as well as their asymptotic distributions.

Using these definitions, we show in the Supplementary Appendix B that
\begin{align}
\widehat{\mathbf{w}} & = \arg \min _{\mathbf{w} \in \mathcal{H}} (\widehat{\mathbf{w}}_{\mathrm{OLS}}-\mathbf{w})^{\prime} (\widetilde{\mathbb{Z}}^{\prime} \widetilde{\mathbb{Z}} / T_0) (\widehat{\mathbf{w}}_{\mathrm{OLS}}-\mathbf{w}) \notag \\
& = \Pi_{\mathcal{H}, T_0} \widehat{\mathbf{w}}_{\mathrm{OLS}}, \label{5.3}
\end{align}
indicating that the constrained estimator is a projection of the unconstrained estimator onto the constrained set $\mathcal{H}$.

The asymptotic distribution of $\sqrt{T_0} (\widehat{\mathbf{w}} - \mathbf{w}_0)$ can be expressed as a projection of the limiting distribution of an unconstrained estimator $\widehat{\mathbf{w}}_{\rm OLS}$ into a convex set, which is the so-called tangent cone of $\mathcal{H}$ evaluated at $\mathbf{w}_0$. A formal definition of the tangent cone, denoted by $T_{\mathcal{H}, \mathbf{w}_0}$, is provided in Supplementary Appendix B.

To Analyze the asymptotic distribution of $\widehat{\alpha}_1$, we first establish the asymptotic distribution of the MF-SCM estimator $\widehat{\mathbf{w}}$. We begin by listing some regularity conditions required for proving the main distribution results.

\begin{assumption} \label{ass6}
{\it The data $\{Y_{j, t}: j=1, \ldots, J_1 \}_{t=1}^{T_0}$, $\{Y_{j, t}: j\in \mathcal{J}_2\}_{t=n\widetilde{m}_j (n \in \mathbb{N}_{+})}$, $\{Y_{j, t-k/m_j}^{(m_j)} : k=0, \ldots, K_j; j\in \mathcal{J}_3 \}_{t=1}^{T_0}$ and $\{X_{j, q, t}: j=1, \ldots, J+1; q=1, \ldots, Q\}_{t=1}^{T_0}$ are weakly dependent stationary process. Under these conditions, standard laws of large numbers apply, which imply that 
$T_0^{-1} \sum_{t=1}^{T_0} \widetilde{\mathbf{Y}}_{t} \xrightarrow{p}$ $\mathbb{E}(\widetilde{\mathbf{Y}}_{t})$ and $\widetilde{\mathbb{Y}}^{\prime} \widetilde{\mathbb{Y}} / T_0 \equiv T_0^{-1} \sum_{t=1}^{T_0} \widetilde{\mathbf{Y}}_{t} \widetilde{\mathbf{Y}}_{t}^{\prime} \xrightarrow{p} \mathbb{E}(\widetilde{\mathbf{Y}}_{t} \widetilde{\mathbf{Y}}_{t}^{\prime})$ for any $t \in \mathcal{T}_0$.
$\mathbb{E}(\widetilde{\mathbf{Y}}_{t} \widetilde{\mathbf{Y}}_{t}^{\prime})$ is positive definite. Let $\phi = \lim _{T_0, T_1 \rightarrow \infty} \sqrt{T_1 / T_0}$. Then, $\phi$ is a finite nonnegative constant.}
\end{assumption}

\begin{assumption} \label{ass7}
{\it $\{u_{\tilde{t}}\}_{\tilde{t}=1}^{T_z}$ has zero mean and satisfies $T_z^{-1/2} \sum_{\tilde{t}=1}^{T_z} \widetilde{\mathbf{Z}}_{\tilde{t}} u_{\tilde{t}} \xrightarrow{d} N (0, \Sigma_1)$, where $\Sigma_1 = \lim _{T_0 \rightarrow \infty} T_z^{-1} \sum_{t=1}^{T_z} \sum_{s=1}^{T_z} \mathbb{E}(u_{t} u_{s} \widetilde{\mathbf{Z}}_{t} \widetilde{\mathbf{Z}}_{s}^{\prime})$.}
\end{assumption}

\begin{assumption} \label{ass8}
{\it Let $v_{1, t} = \alpha_{1, t} - \alpha_1 + u_{t}$. We assume that $v_{1,t}$ has zero mean and satisfies a central limit theorem: $T_1^{-1/2} \sum_{t=T_0+1}^T v_{1, t} \xrightarrow{d} N (0, \Sigma_v)$, where $\Sigma_v = \lim_{T_1 \rightarrow \infty} T_1^{-1}$ $\sum_{t=T_0+1}^T \sum_{s=T_0+1}^T \mathbb{E} (v_{1, t} v_{1, s})$.}
\end{assumption}

\begin{assumption} \label{ass9}
{\it Let $\boldsymbol{\omega}_t = (Y_{1, t}, \widetilde{\boldsymbol{Y}}_t^{\prime}, \alpha_{1, t} D_{1,t})$ for $t=1, \ldots, T$. Assume that $\{\alpha_{1, t} D_{1,t}\}_{t=T_0+1}^T$ is a weakly dependent stationary process. In conjunction with Assumption \ref{ass6}, it follows that both $\{\boldsymbol{\omega}_t\}_{t=1}^{T_0}$ and $\{\boldsymbol{\omega}_t\}_{t=T_0+1}^T$ are weakly dependent stationary processes. Define $\rho(\nu) = \max _{1 \leq t \leq T} \max _{1 \leq i, j \leq J+1}$ $|\operatorname{cov} (\omega_{i, t}, \omega_{j, t+\nu}) / \sqrt{\operatorname{var} (\omega_{i, t}) \operatorname{var} (\omega_{j, t+\nu} )} |$, where $\omega_{i, t}$ is the $i$-th component of $\boldsymbol{\omega}_t$, $i=1,\ldots,J+2$. Then there exist finite constant $0<\lambda<1$ such that $\rho(\nu)= O(\lambda^\nu)$.}
\end{assumption}

Assumption \ref{ass6} requires that the outcome sequences $\{Y_{j, t}: j\in \mathcal{J}_1\}_{t=1}^{T_0}$, $\{Y_{j, t}: j\in \mathcal{J}_2\}_{t=n\widetilde{m}_j (n \in \mathbb{N}_{+})}$, and $\{Y_{j, t-k/m_j}^{(m_j)} : k=0, \ldots, K_j; j\in \mathcal{J}_3 \}_{t=1}^{T_0}$, as well as the covariate sequences $\{X_{j, q, t}: j=1, \ldots, J+1; q=1, \ldots, Q\}_{t=1}^{T_0}$ are weakly dependent stationary processes, which ensure that the laws of large numbers hold for both the first-order and second-order moments of the constructed sequence $\widetilde{\mathbf{Y}}_{t}$. 
Assumption \ref{ass7} imposes a central limit theorem-type condition on the error term $u_{\tilde{t}}$. Similar assumptions to Assumptions \ref{ass6} and \ref{ass7} are also imposed in \citet{li2020statistical}. 
Under Assumptions \ref{ass6} and \ref{ass7}, we have $\sqrt{T_0} (\widehat{\mathbf{w}}_{\rm OLS} - \mathbf{w}_0) \xrightarrow{d}$ $\mathrm{Normal} (0, A^{-1} \Sigma_1 A^{-1})$, where $A = \mathbb{E}(\widetilde{\mathbf{Y}}_{t} \widetilde{\mathbf{Y}}_{t}^{\prime})$. Assumption \ref{ass8} requires that a central limit theorem applies to a partial sum of $v_{1, t}$. Assumption \ref{ass9} is also used in \citet{li2017estimation} and requires that the data is a weakly dependent stationary process with an exponential decay rate.

Let $G_1$ denote the limiting (normal) distribution of $\sqrt{T_0} (\widehat{\mathbf{w}}_{\rm OLS} - \mathbf{w}_0)$, i.e., $G_1 \sim \mathrm{Normal} (0, A^{-1} \Sigma_1 A^{-1})$.

\begin{theorem} \label{th2}
{\it Under Assumptions \ref{ass6}-\ref{ass9}, we have
\begin{eqnarray}
\sqrt{T_0} (\widehat{\mathbf{w}} - \mathbf{w}_0) \xrightarrow{d} \Pi_{T_{\mathcal{H}, \mathbf{w}_0}} G_1. \label{5.4}
\end{eqnarray}
}
\end{theorem}

Theorem \ref{th2} states that the limiting distribution of the constrained estimator can be represented as a projection of the limiting distribution of the unconstrained (least squares) estimator onto the tangent cone $T_{\mathcal{H}, \mathbf{w}_0}$. There is a simple interpretation of the above result. The range of $G_1$ is $\mathcal{R}^J$. 
However, when the constraints are binding, the range of $\sqrt{T_0} (\widehat{\mathbf{w}} - \mathbf{w}_0)$ is a convex subset of $\mathcal{R}^J$. It can be shown that the asymptotic (as $T_0 \rightarrow \infty$) range of $\sqrt{T_0} (\widehat{\mathbf{w}} - \mathbf{w}_0)$ is exactly $T_{\mathcal{H}, \mathbf{w}_0}$, the tangent cone of $\mathcal{H}$ at $\mathbf{w}_0$.
Therefore, the asymptotic distribution of $\sqrt{T_0} (\widehat{\mathbf{w}} - \mathbf{w}_0)$ is the projection of the limiting distribution of $\sqrt{T_0} (\widehat{\mathbf{w}}_{\rm OLS} - \mathbf{w}_0)$ onto the asymptotic range of $\sqrt{T_0} (\widehat{\mathbf{w}} - \mathbf{w}_0)$.

With the help of Theorem \ref{th2}, we derive the asymptotic distribution of $\sqrt{T_1} (\widehat{\alpha}_1 - \alpha_1)$ as follows.

\begin{theorem} \label{th3}
{\it Under Assumptions \ref{ass6}-\ref{ass9}, we have
\begin{eqnarray}
\sqrt{T_1} (\widehat{\alpha}_1 - \alpha_1) \xrightarrow{d} -\phi \mathbb{E} (\widetilde{\boldsymbol{Y}}_t^{\prime}) \Pi_{T_{\mathcal{H}, \mathbf{w}_0}} G_1 + G_2, \label{5.5}
\end{eqnarray}
where $G_2$ is independent of $G_1$ and follows a normal distribution $\mathrm{Normal} (0, \Sigma_v )$, and $\Sigma_v = \lim _{T_1 \rightarrow \infty} T_1^{-1} \sum_{t=T_0+1}^T \sum_{s=T_0+1}^T \mathbb{E} (v_{1, t} v_{1, s} )$.}
\end{theorem}

Theorem \ref{th3} establishes the asymptotic distribution of $\sqrt{T_1} (\widehat{\alpha}_1 - \alpha_1)$, which consists of two components. The first component reflects the dependence of Theorem \ref{th3} on Theorem \ref{th2}. The second component, $G_2$, captures the post-treatment disturbances and converges to a mean-zero normal distribution with variance $\Sigma_v$.

\section{Inference Theory} \label{sec:5}
In this section, we discuss inference methods for the ATE estimator $\widehat{\alpha}_1$.
Although projection theory can be employed to describe the asymptotic distribution of $\sqrt{T_1} (\widehat{\alpha}_1 - \alpha_1)$, its practical implementation is challenging, as it requires knowledge of $\mathbf{w}_0$ to calculate $T_{\mathcal{H}, \mathbf{w}_0}$. We demonstrate in this section that a block subsampling method, introduced by \citet{politis1999subsampling}, provides a valid alternative for inference.
Notably, this method does not need to know $\mathbf{w}_0$.

The asymptotic distribution of the MF-SCM coefficient estimators depends on whether the true parameters lie on the boundary of the parameter space. In practice, it is often unclear which constraints are binding, complicating the application of asymptotic theory for inference. Moreover, when parameters fall on the boundary, the standard bootstrap method fails (\citeauthor{andrews2000inconsistency}, \citeyear{andrews2000inconsistency}; \citeauthor{fang2019inference}, \citeyear{fang2019inference}).
To address this issue, we follow \citet{li2020statistical} and adopt a computationally convenient subsampling method. This method is robust to whether constraints are binding, partially binding, or nonbinding. That is, it adapts to the constraint structure without requiring prior knowledge of which constraints, if any, are binding, nor does it necessitate identifying the coefficients subject to binding constraints. However, because this method does not take the order of the original sequence into account, it is generally unsuitable for time series data. For collecting the subsamples, we employ a block subsampling approach, which offers greater robustness to temporal dependence in the outcomes.

For $t \in \mathcal{T}_0 \cup \mathcal{T}_1$, we assume that the outcome $Y_{1, t}$ follows the model
\begin{align}
Y_{1, t} = \widetilde{\boldsymbol{Y}}_t^{\prime} \mathbf{w}_0 + D_{1,t} \alpha_{1, t} + u_{t}. \label{6.1}
\end{align}
It should be noted that the constraint $\mathbf{w}_0 \in \mathcal{H}$ may lead to cases where assumption (\ref{6.1}) does not hold exactly, implying a potential misspecification in the relationship between $Y_{1, t}$ and $\widetilde{\boldsymbol{Y}}_t^{\prime}$.
When assumption (\ref{6.1}) holds, we can derive
\begin{align}
E & \stackrel{\text{def}} = \sqrt{T_1} (\widehat{\alpha}_1 - \alpha_1) \notag \\
& = \sqrt{T_1} \left(\frac{1}{T_1} \sum_{t=T_0+1}^T \widehat{\alpha}_{1,t} - \alpha_1 \right) \notag \\
& = \sqrt{T_1} \left(\frac{1}{T_1} \sum_{t=T_0+1}^T (\widetilde{\boldsymbol{Y}}_t^{\prime} \mathbf{w}_0 + \alpha_{1, t} + u_{t} - \widetilde{\boldsymbol{Y}}_t^{\prime} \widehat{\mathbf{w}}) - \alpha_1 \right) \notag \\
& = -\sqrt{\frac{T_1}{T_0}} \left[\frac{1}{T_1} \sum_{t=T_0+1}^T \widetilde{\boldsymbol{Y}}_t^{\prime}\right] \sqrt{T_0} (\widehat{\mathbf{w}} - \mathbf{w}_0) + \frac{1}{\sqrt{T_1}} \sum_{t=T_0+1}^T v_{1, t} \notag \\
& \equiv E_1 + E_2, \label{6.2}
\end{align}
where $E_1 = -\sqrt{\frac{T_1}{T_0}} \left[\frac{1}{T_1} \sum_{t=T_0+1}^T \widetilde{\boldsymbol{Y}}_t^{\prime}\right] \sqrt{T_0} (\widehat{\mathbf{w}} - \mathbf{w}_0)$, and $E_2 = \frac{1}{\sqrt{T_1}} \sum_{t=T_0+1}^T v_{1, t}$.

Expression (\ref{6.2}) decomposes $\widehat{\alpha}_1$ into two terms: the first term is associated with $\widehat{\mathbf{w}}$, while the second term is unrelated to $\widehat{\mathbf{w}}$ but depends on $T_1$. A direct application of the block subsampling method is generally ineffective. Instead, the appropriate approach involves applying the block subsampling method exclusively to the $\sqrt{T_0} (\widehat{\mathbf{w}} - \mathbf{w}_0)$ term, while employing the bootstrap method for the second term that is unrelated to $\widehat{\mathbf{w}}$.
To simplify the block subsampling method, we impose an additional assumption that $u_{t}$ and $v_{1, t}$ are both serially uncorrelated, ensuring a straightforward implementation.

Since $\sqrt{T_0} (\widehat{\mathbf{w}} - \mathbf{w}_0)$ is the only term related to the constrained estimator, (\ref{6.2}) suggests that the block subsampling method should be applied exclusively to this term.
Below, we outline the block subsampling steps. In Supplementary Appendix C, we demonstrate that when $v_{1, t}$ is serially uncorrelated, the covariance $\Sigma_v$ can be consistently estimated as 
$$\widehat{\Sigma}_v = T_1^{-1} \sum_{t=T_0+1}^T \widehat{v}_{1, t}^2,$$
where $\widehat{v}_{1, t} = \widehat{\alpha}_{1, t} - \widehat{\alpha}_1$. We then generate $v_{1, t}^* \sim$ iid $N(0, \widehat{\Sigma}_v)$ for $t=T_0+1, \ldots, T$. Let $m$ denote the block subsample size, satisfying $m \rightarrow \infty$ and $m/T_0 \rightarrow 0$ as $T_0 \rightarrow \infty$. 
We begin by conducting block subsampling on the matrix $\widetilde{\mathbb{Y}}$. Specifically, for each $b = 1, \ldots, T_0 - m + 1$, we construct an outcome block $\{\widetilde{\mathbf{Y}}_{1}^*, \ldots, \widetilde{\mathbf{Y}}_{m}^*\} = \{\widetilde{\mathbf{Y}}_{b}, \ldots, \widetilde{\mathbf{Y}}_{b+m-1}\}$. Thus, the first block is $\{\widetilde{\mathbf{Y}}_{1}, \widetilde{\mathbf{Y}}_{2}, \ldots, \widetilde{\mathbf{Y}}_{m}\}$ and the last is $\{\widetilde{\mathbf{Y}}_{T_0-m+1}, \widetilde{\mathbf{Y}}_{T_0-m+2}, \ldots, \widetilde{\mathbf{Y}}_{T_0}\}$. In total, there are $T_0-m+1$ such overlapping blocks. Each subsampled block of $\widetilde{\mathbb{Y}}$, consisting of $m$ consecutive rows.
Let $\{Y_{1, 1}^*, \ldots, Y_{1, m}^*\}$ denote the corresponding outcome subsamples of the treated unit in the $b$-th block.
Then, we can estimate $\mathbf{w}_0$ by the constrained least squares method, that is,
$$\widehat{\mathbf{w}}_m^* = \arg\min_{\mathbf{w} \in \mathcal{H}} \sum_{t=1}^m (Y_{1, t}^* - \widetilde{\mathbf{Y}}_t^{* \prime} \mathbf{w})^2.$$
The subsampling-bootstrap version of $E$ is given by
\begin{align}
\widehat{E}^* = & -\sqrt{\frac{T_1}{T_0}} \left[\frac{1}{T_1} \sum_{t=T_0+1}^T \widetilde{\boldsymbol{Y}}_t^{\prime}\right] \sqrt{m} (\widehat{\mathbf{w}}_m^* - \widehat{\mathbf{w}}) \notag \\
& + \frac{1}{\sqrt{T_1}} \sum_{t=T_0+1}^T v_{1, t}^*. \label{6.3}
\end{align}

We repeat the above process a large number of times ($N$ times) to generate a set of subsampling-bootstrap statistics, denoted as $\{\widehat{E}_n^*\}_{n=1}^N$. Based on these statistics, we can construct confidence intervals for $\alpha_1$. Specifically, we sort these statistics in ascending order, i.e., $\widehat{E}_{(1)}^* \leq \widehat{E}_{(2)}^* \leq \ldots \leq \widehat{E}_{(N)}^*$, and obtain the $1-\alpha$ confidence interval for $\alpha_1$ as
$$
[\widehat{\alpha}_1 - T_1^{-1/2} \widehat{E}_{((1-\alpha/2)N)}^*, ~\widehat{\alpha}_1 - T_1^{-1/2} \widehat{E}_{((\alpha/2)N)}^*].
$$

\begin{theorem} \label{th4}
{\it Under Assumptions \ref{ass6}-\ref{ass9} and the assumptions that $u_{t}$ and $v_{1, t}$ are serially uncorrelated, and that $m \rightarrow \infty$ and $m/T_0 \rightarrow 0$ as $T_0 \rightarrow \infty$, the $(1-\alpha)$ confidence interval of $\alpha_1$ can be consistently estimated by 
$[\widehat{\alpha}_1 - T_1^{-1/2} \widehat{E}_{((1-\alpha/2)N)}^*, ~\widehat{\alpha}_1 - T_1^{-1/2} \widehat{E}_{((\alpha/2)N)}^*]$
for any $\alpha \in (0,1)$, i.e.,
$$\mathbb{P}(\alpha_1 \in [\widehat{\alpha}_1 - T_1^{-1/2} \widehat{E}_{((1-\alpha/2)N)}^*, ~\widehat{\alpha}_1 - T_1^{-1/2} \widehat{E}_{((\alpha/2)N)}^*]) \rightarrow 1-\alpha, ~as~ T_0 \rightarrow \infty.$$
}
\end{theorem}

Theorem \ref{th4} demonstrates that the above subsampling-bootstrap method consistently estimates the confidence intervals for $\alpha_1$, and provides a theoretical justification for this method, reinforcing its applicability in empirical settings.

In summary, the block-subsampling inference procedure for the ATE is summarized in Algorithm \ref{alg:mfscm-inference}.

\begin{algorithm}[htb]
\caption{Block-Subsampling Inference for ATE in MF-SCM }
\label{alg:mfscm-inference}
\begin{algorithmic}[1]
\STATE \textbf{Input:} Estimated $(\widehat{\mathbf w},\widehat{\boldsymbol\theta})$, aligned pre/post outcomes, significance level $\alpha$, and bootstrap size $N$.
\STATE Compute $\widehat{\alpha}_{1,t}$ and $\widehat{\alpha}_1$, then estimate $\widehat{\Sigma}_v=T_1^{-1}\sum_{t=T_0+1}^{T}\widehat{v}_{1,t}^{\,2}$.
\STATE Choose block size $m$ such that $m_b\to\infty$ and $m/T_0\to0$ (empirically $m=\lfloor T_0^{0.8}\rfloor$).
\STATE Construct all overlapping pre-treatment blocks of length $m_b$ from $\widetilde{\mathbb Y}$ and corresponding treated outcomes.
\FOR{$n=1,\ldots,N$}
\STATE Draw a block start index $b$ uniformly at random from \(\{1,\ldots,T_0-m+1\}\) (with replacement), i.e., \(b_n \sim \mathrm{Unif}\{1,\ldots,T_0-m+1\}\). 
\STATE Construct the outcome block \(\{\widetilde{\mathbf{Y}}_{t}^{*}\}_{t=1}^{m}=\{\widetilde{\mathbf{Y}}_{b+t-1}\}_{t=1}^{m}\) and the treated-unit outcome block \(\{Y_{1,t}^{*}\}_{t=1}^{m}=\{Y_{1,b+t-1}\}_{t=1}^{m}\).
\STATE Estimate the constrained least squares estimator on the block
$$\widehat{\mathbf{w}}_m^* = \arg\min_{\mathbf{w} \in \mathcal{H}} \sum_{t=1}^m (Y_{1, t}^* - \widetilde{\mathbf{Y}}_t^{* \prime} \mathbf{w})^2.$$
\STATE Draw $v_{1,t}^{*}\sim iid\ N(0,\widehat{\Sigma}_v)$ for $t=T_0+1,\ldots,T$.
\STATE Form the subsampling-bootstrap statistic $\widehat{E}_n^*$ using (\ref{6.3}).
\ENDFOR
\STATE Sort $\{\widehat{E}_n^*\}_{n=1}^N$ in ascending order: $\widehat{E}_{(1)}^* \leq \widehat{E}_{(2)}^* \leq \ldots \leq \widehat{E}_{(N)}^*$
\STATE Construct the $(1-\alpha)$ confidence interval:
\[
\left[\widehat{\alpha}_1 - T_1^{-1/2}\widehat{E}_{((1-\alpha/2)N)}^*,\ 
\widehat{\alpha}_1 - T_1^{-1/2}\widehat{E}_{((\alpha/2)N)}^* \right].
\]
\STATE \textbf{Output:} Confidence interval for $\alpha_1$.
\end{algorithmic}
\end{algorithm}

\section{Simulation}\label{sec:6}
In this section, we conduct simulation experiments to examine the asymptotic optimality of MF-SCM estimator in Theorem \ref{th1} and to validate the effectiveness of the confidence intervals proposed in Section \ref{sec:5}.
In both simulation experiments, the covariates available to the researcher are observed only at the baseline frequency. Specifically, for each control unit \(j=2,\ldots,J+1\) and baseline period \(t=1,\ldots,T_0+T_1\), we generate the observed baseline-frequency covariates as
\[
\mathbf{X}_{j,t} \sim \mathcal{N}(\boldsymbol{\varphi}_j,\mathbf{I}_Q),
\qquad
\boldsymbol{\varphi}_j=(\varphi_{j,1},\ldots,\varphi_{j,Q})^\prime,
\qquad
\varphi_{j,q}\sim U(-3,3),
\]
where \(\mathbf{I}_Q\) denotes the \(Q\times Q\) identity matrix. For a given number of control units \(J\), we define \(J_1=\lfloor J/3 \rfloor\) and \(J_2=\lfloor 2J/3 \rfloor\), and let \(\mathcal{J}_1=\{2,\ldots,J_1+1\}\), \(\mathcal{J}_2=\{J_1+2,\ldots,J_2+1\}\), and \(\mathcal{J}_3=\{J_2+2,\ldots,J+1\}\) denote the baseline-frequency, low-frequency, and high-frequency control units, respectively. For each control unit in \(\mathcal{J}_1\cup\mathcal{J}_2\), the latent baseline-frequency outcome follows
\begin{align}  \label{eq:6.1}
Y_{j,t}^N = \alpha_0 + \sum_{p=0}^{P-1} \boldsymbol{\beta}_{p}^{\prime} \mathbf{X}_{j,t-p} + \epsilon_{j,t},
\qquad t=1,\ldots,T_0+T_1,
\end{align}
where \(\alpha_0=0\), \( \boldsymbol{\beta}_{p} \sim [\mathcal{U}(-1,1)]^Q \), and \( \epsilon_{j,t} \sim \mathcal{N}(0,\sigma^2) \). The covariate processes are initialized for the required lagged periods. Throughout the simulations, we set \(Q=3\), \(P=2\), and \(\sigma=1\).

For controls in \(\mathcal{J}_3\), we additionally generate an unobserved high-frequency covariate process
\[
\mathbf{X}_{j,t-(k-1)/m_j}^{(m_j)} \sim \mathcal{N}(\boldsymbol{\varphi}^{(h)}_j,\mathbf{I}_Q),
\qquad
\boldsymbol{\varphi}^{(h)}_j=(\varphi^{(h)}_{j,1},\ldots,\varphi^{(h)}_{j,Q})^\prime,
\qquad
\varphi^{(h)}_{j,q}\sim U(-3,3),
\]
for \(t=1,\ldots,T_0+T_1\) and \(k=1,\ldots,m_j\). We do not impose any deterministic relation between the observed baseline-frequency covariates \(\mathbf{X}_{j,t}\) and the latent high-frequency covariates \(\mathbf{X}_{j,t-(k-1)/m_j}^{(m_j)}\). The untreated high-frequency outcomes satisfy the high-frequency distributed lag model in \eqref{4.1h}:
\begin{equation}
Y_{j,t-(k-1)/m_j}^{(m_j),N}
=
\alpha_0^{(h)}
+
\sum_{p=0}^{P-1}
\boldsymbol{\beta}_{p}^{(h)\prime}
\mathbf{X}_{j,t-(k-1)/m_j-p/m_j}^{(m_j)}
+
\epsilon_{j,t-(k-1)/m_j}^{(h)},
\label{eq:sim-high-dgp}
\end{equation}
for \(j\in\mathcal{J}_3\), \(t=1,\ldots,T_0+T_1\), and \(k=1,\ldots,m_j\), where \(\alpha_0^{(h)}=0\), \(\boldsymbol{\beta}_{p}^{(h)} \sim [\mathcal{U}(-1,1)]^Q\), and \(\epsilon_{j,t-(k-1)/m_j}^{(h)}\sim\mathcal{N}(0,\sigma_h^2)\). We set \(m_j=K_0=3\) and \(\sigma_h=1\).

The control units are partitioned into three groups according to their sampling frequencies. For \(j\in\mathcal{J}_1\), the baseline-frequency outcome is observed directly, so \(Y_{j,t}=Y_{j,t}^N\). For \(j\in\mathcal{J}_2\), only a low-frequency aggregate is observed:
\begin{equation}
Y_{j,t^{(1:\widetilde m_j)}} = \sum_{s=1}^{\widetilde m_j} W_{j,s} Y_{j,t-\widetilde m_j+s}^{N},
\qquad
t \in \{\widetilde m_j, 2\widetilde m_j, \ldots, \lfloor T/\widetilde m_j \rfloor \widetilde m_j\},
\qquad
\sum_{s=1}^{\widetilde m_j} W_{j,s}=1,
\label{eq:sim-low-obs}
\end{equation}
where the latent baseline-frequency path \(\{Y_{j,t}^{N}\}\) is not observed. In the simulations we take \(\widetilde m_j=4\) and \(W_{j,s}=1/\widetilde m_j\). For \(j\in\mathcal{J}_3\), the observed series is the high-frequency outcome path \(\{Y_{j,t-(k-1)/m_j}^{(m_j),N}\}\) generated by \eqref{eq:sim-high-dgp}.

After generating the control units, we construct the untreated path of the treated unit using an oracle unit-weight vector \(\boldsymbol{\omega}_0\) and oracle high-frequency MIDAS coefficients \((\boldsymbol{\zeta}_j^0)_{j\in\mathcal{J}_3}\). Define
\[
\widetilde{Y}_{j,t}^{N,0}
=
\begin{cases}
Y_{j,t}^{N}, & j\in\mathcal{J}_1\cup\mathcal{J}_2,\\
\sum_{k=1}^{K_0} B_j(k;\boldsymbol{\zeta}_j^0)Y_{j,t-(k-1)/m_j}^{(m_j),N}, & j\in\mathcal{J}_3,
\end{cases}
\]
where \(B_j(\cdot;\boldsymbol{\zeta}_j^0)\) denotes the oracle high-frequency MIDAS weight vector for unit \(j\). The baseline-frequency covariates of the treated unit are defined by
\[
\mathbf{X}_{1,t}
=
\sum_{j=2}^{J+1}\omega_{0,j}\mathbf{X}_{j,t},
\]
and the untreated treated-unit outcome is generated by
\[
Y_{1,t}^{N}
=
\sum_{j=2}^{J+1}\omega_{0,j}\widetilde{Y}_{j,t}^{N,0}
+u_t,
\qquad
u_t\sim\mathcal{N}(0,\sigma_u^2).
\]
In both simulation experiments, we set \(\sigma_u=0.5\). To ensure that the simulation results are comparable across different values of \(T_0\) and \(T_1\), we keep the oracle components fixed throughout each simulation design. 
Specifically, \(\boldsymbol{\omega}_0\) and \((\boldsymbol{\zeta}_j^0)_{j\in\mathcal{J}_3}\) are drawn once and then kept fixed across the simulation grid. Moreover, \(\boldsymbol{\omega}_0\) is generated under a balanced group-mass rule: one third of the total unit-weight mass is assigned to each of the baseline-frequency, low-frequency, and high-frequency groups, while the within-group unit weights are obtained from one softmax draw per group.

Accordingly, the aligned untreated control-unit outcome entering MF-SCM is
\[
\widetilde{Y}_{j,t}^{N}(\boldsymbol{\theta})=
\begin{cases}
Y_{j,t}^{N}, & j\in\mathcal{J}_1,\\
\widehat{Y}_{j,t}, & j\in\mathcal{J}_2,\\
\sum_{k=1}^{K_0} B_j(k;\boldsymbol{\zeta}_j)Y_{j,t-(k-1)/m_j}^{(m_j),N}, & j\in\mathcal{J}_3,
\end{cases}
\]
where \(\widehat{Y}_{j,t}\) denotes the reconstructed baseline-frequency outcome obtained from the low-frequency reconstruction model estimated on the pre-treatment sample, and \(B_j(k;\boldsymbol{\zeta}_j)\) is parameterized by the dictionary specification in \eqref{eq:dict-midas}. No frequency alignment is applied to the observed covariates \(\mathbf{X}_{j,t}\), since the estimator only uses the baseline-frequency covariates. For high-frequency units, the latent high-frequency covariates \(\mathbf{X}_{j,t-(k-1)/m_j}^{(m_j)}\) affect the DGP of \(Y_{j,t-(k-1)/m_j}^{(m_j),N}\), but they are not observed by the estimator.

\subsection{Validation of Asymptotic Optimality} \label{sec:6-1}
We first examine the asymptotic optimality result in Theorem \ref{th1}.
We vary the number of pre-treatment periods over 
\(T_0 \in \{20,40,80,160,320,640,1280,2560\}\), while fixing \(T_1=100\). To validate the asymptotic optimality result, we evaluate
\begin{equation}
\rho(T_0)
:=
\frac{R_{T_1}(\widehat{\mathbf{w}},\widehat{\boldsymbol{\theta}})}
{\inf_{\mathbf{w}\in\mathcal{H},\boldsymbol{\theta} \in \mathcal{R}^{M}}R_{T_1}(\mathbf{w},\boldsymbol{\theta})}
\label{eq:sim-risk-ratio}
\end{equation}
where
\begin{equation}
R_{T_1}(\mathbf{w},\boldsymbol{\theta})
=
\frac{1}{T_1}\sum_{t\in\mathcal{T}_1}
\mathbb{E}\left(
Y_{1,t}^{N}
-
\sum_{j=2}^{J+1}w_j\widetilde{Y}_{j,t}^{N}(\boldsymbol{\theta})
\right)^2.
\label{eq:sim-risk}
\end{equation}

Because the risk in \eqref{eq:sim-risk} coincides with the definition in Section~\ref{sec:3}, namely
\(
R_{T_1}(\mathbf{w},\boldsymbol{\theta})=\mathbb{E}L_{T_1}(\mathbf{w},\boldsymbol{\theta}),
\)
the expectation is taken over all randomness entering the post-treatment loss. In the implementation, this deterministic risk is rewritten in the lifted convex form induced by the dictionary representation, approximated by a pooled Monte Carlo quadratic program on a common evaluation surface, and then evaluated at estimators obtained from independent training panels. The full construction of this Monte Carlo approximation is given in Appendix H.

We compare three estimators on this common risk surface: MF-SCM, MF-SCM w/o MIDAS, and Baseline-Freq Only. The proposed MF-SCM uses the full feasible class of unit weights and MIDAS weights. MF-SCM w/o MIDAS restricts each high-frequency unit to equal aggregation weights,
\[
B_j(1;\boldsymbol{\zeta}_j)=B_j(2;\boldsymbol{\zeta}_j)=B_j(3;\boldsymbol{\zeta}_j)=\frac{1}{3},
\qquad j\in\mathcal J_3,
\]
while Baseline-Freq Only imposes \(w_j=0\) for all \(j\in\mathcal J_2\cup\mathcal J_3\) and fits the synthetic control using only the baseline-frequency units. 

\begin{figure}[htb]
    \centering
    \includegraphics[width=\textwidth]{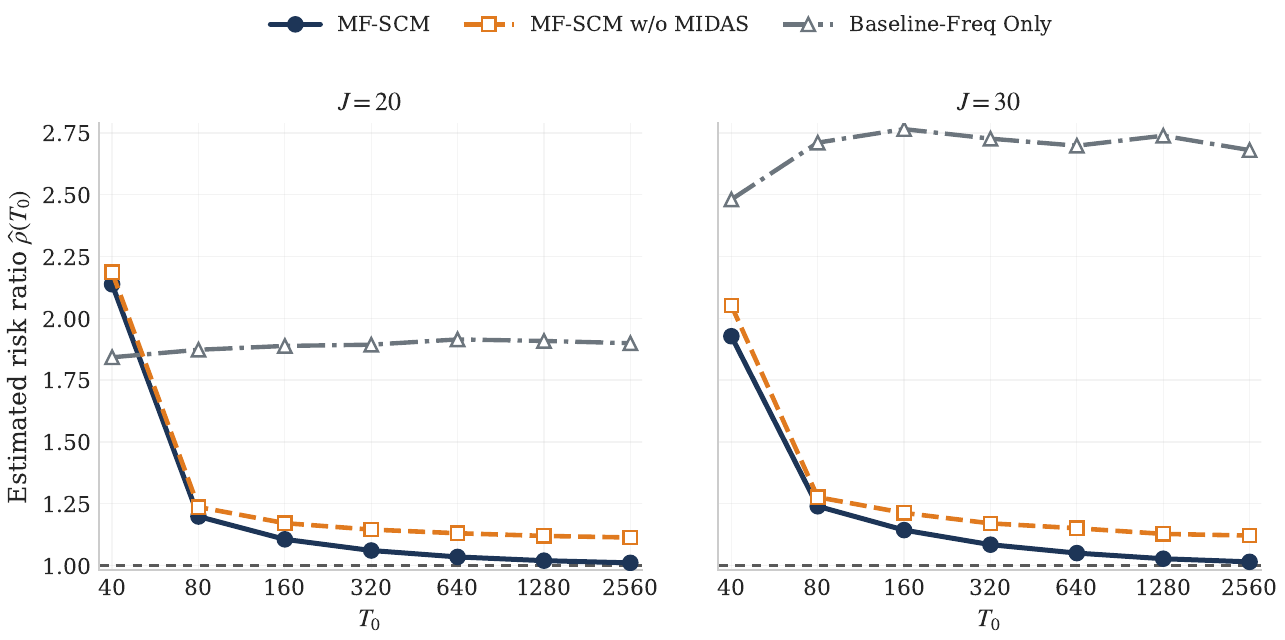}
    \caption{Risk-ratio comparison under the mixed-frequency DGP described in Section \ref{sec:6}, with balanced group-level unit weights, fixed front-loaded oracle MIDAS weights, \(T_1=100\), \(S=500\), and \(M=1000\). The left and right panels report \(J=20\) and \(J=30\), respectively.}
    \label{fig:sim1}
\end{figure}

Figure \ref{fig:sim1} reports the average risk ratio for \(J=20\) and \(J=30\). The MF-SCM risk ratio decreases toward one as \(T_0\) increases, consistent with the asymptotic optimality result in Theorem \ref{th1}. 
By contrast, MF-SCM without MIDAS remains bounded away from one, and Baseline-Frequency Only performs substantially worse. 
These patterns indicate that both frequency alignment and flexible MIDAS aggregation are important for achieving the oracle performance predicted by the theory.

\subsection{Validation of the Effectiveness of Confidence Intervals}\label{sec:6-2}
This simulation uses the same control-unit DGP and the same oracle treated-unit construction introduced above, but now focuses on interval coverage. We set \(J=20\), adopt the same low-frequency latent aggregation model in \eqref{eq:aggregation} together with the baseline-frequency reconstruction procedure implied by \eqref{eq:latent_model}--\eqref{eq:combined_model} and Appendix G, vary \(T_0\in\{40,80,160,320,640\}\) and \(T_1\in\{20,40,80,160\}\), and keep one fixed oracle unit-weight vector together with one fixed collection of nonnegative, front-loaded oracle MIDAS coefficients across the full \((T_0,T_1)\) grid. The unit-weight vector satisfies the same balanced group-mass rule as above, so each frequency group carries one third of the total unit-weight mass while the within-group unit weights remain fixed throughout the experiment.

For each \((T_0,T_1)\) design point, the untreated treated-unit outcome follows the fixed oracle specification described above. We run 5000 Monte Carlo repetitions and construct confidence intervals using 1000 subsampling-bootstrap iterations, with block length \(m_{T_0}=\max\{10,\lfloor T_0^{0.5}\rfloor\}\). Table \ref{tab:1} reports the Monte Carlo coverage rates and average interval lengths for nominal levels \(90\%\), \(95\%\), and \(99\%\).

The coverage rates are close to the nominal levels overall, though the smallest-sample settings exhibit mild under-coverage. As \(T_1\) increases, the confidence intervals become shorter and the coverage rates become more stable, while larger \(T_0\) further improves precision.

\begin{table}[htbp]
    \centering
    \caption{Monte Carlo coverage rates and average interval lengths for \(J=20\) with varying \(T_0\) and \(T_1\).} 
    \label{tab:1}
    \begin{tabular}{c|cc|cc|cc}
        \toprule
        \multicolumn{7}{c}{\(T_1=20\)} \\
        \midrule
        \multirow{2}{*}{\(T_0\)} & \multicolumn{2}{c|}{\(\alpha = 0.1\)} & \multicolumn{2}{c|}{\(\alpha = 0.05\)} & \multicolumn{2}{c}{\(\alpha = 0.01\)} \\
        \cmidrule(lr){2-3} \cmidrule(lr){4-5} \cmidrule(lr){6-7}
         & Coverage & CI Length & Coverage & CI Length & Coverage & CI Length \\
        \midrule
        40  & 0.880 & 0.95 & 0.930 & 1.13 & 0.980 & 1.50 \\
        80  & 0.882 & 0.89 & 0.935 & 1.06 & 0.981 & 1.41 \\
        160 & 0.890 & 0.85 & 0.939 & 1.01 & 0.985 & 1.34 \\
        320 & 0.885 & 0.83 & 0.937 & 0.99 & 0.980 & 1.31 \\
        640 & 0.879 & 0.81 & 0.936 & 0.97 & 0.982 & 1.29 \\
        \midrule
        \multicolumn{7}{c}{\(T_1=40\)} \\
        \midrule
        \multirow{2}{*}{\(T_0\)} & \multicolumn{2}{c|}{\(\alpha = 0.1\)} & \multicolumn{2}{c|}{\(\alpha = 0.05\)} & \multicolumn{2}{c}{\(\alpha = 0.01\)} \\
        \cmidrule(lr){2-3} \cmidrule(lr){4-5} \cmidrule(lr){6-7}
         & Coverage & CI Length & Coverage & CI Length & Coverage & CI Length \\
        \midrule
        40  & 0.881 & 0.73 & 0.937 & 0.87 & 0.983 & 1.15 \\
        80  & 0.903 & 0.67 & 0.952 & 0.80 & 0.990 & 1.06 \\
        160 & 0.896 & 0.63 & 0.946 & 0.75 & 0.985 & 0.99 \\
        320 & 0.885 & 0.60 & 0.939 & 0.72 & 0.986 & 0.95 \\
        640 & 0.895 & 0.59 & 0.943 & 0.70 & 0.987 & 0.93 \\
        \midrule
        \multicolumn{7}{c}{\(T_1=80\)} \\
        \midrule
        \multirow{2}{*}{\(T_0\)} & \multicolumn{2}{c|}{\(\alpha = 0.1\)} & \multicolumn{2}{c|}{\(\alpha = 0.05\)} & \multicolumn{2}{c}{\(\alpha = 0.01\)} \\
        \cmidrule(lr){2-3} \cmidrule(lr){4-5} \cmidrule(lr){6-7}
         & Coverage & CI Length & Coverage & CI Length & Coverage & CI Length \\
        \midrule
        40  & 0.880 & 0.58 & 0.929 & 0.68 & 0.980 & 0.90 \\
        80  & 0.908 & 0.51 & 0.953 & 0.61 & 0.991 & 0.81 \\
        160 & 0.911 & 0.47 & 0.960 & 0.56 & 0.990 & 0.74 \\
        320 & 0.897 & 0.44 & 0.948 & 0.52 & 0.989 & 0.69 \\
        640 & 0.899 & 0.42 & 0.946 & 0.51 & 0.990 & 0.67 \\
        \midrule
        \multicolumn{7}{c}{\(T_1=160\)} \\
        \midrule
        \multirow{2}{*}{\(T_0\)} & \multicolumn{2}{c|}{\(\alpha = 0.1\)} & \multicolumn{2}{c|}{\(\alpha = 0.05\)} & \multicolumn{2}{c}{\(\alpha = 0.01\)} \\
        \cmidrule(lr){2-3} \cmidrule(lr){4-5} \cmidrule(lr){6-7}
         & Coverage & CI Length & Coverage & CI Length & Coverage & CI Length \\
        \midrule
        40  & 0.881 & 0.48 & 0.931 & 0.57 & 0.981 & 0.73 \\
        80  & 0.920 & 0.41 & 0.959 & 0.49 & 0.993 & 0.65 \\
        160 & 0.913 & 0.36 & 0.961 & 0.43 & 0.992 & 0.57 \\
        320 & 0.903 & 0.32 & 0.953 & 0.39 & 0.992 & 0.51 \\
        640 & 0.913 & 0.31 & 0.954 & 0.37 & 0.991 & 0.49 \\
        \bottomrule
    \end{tabular}
\end{table}

\section{Real Data}\label{sec:7}
\subsection{Tax Cuts and Jobs Act of 2017}\label{sec:7-1}
% 加拿大gdp数据 https://www150.statcan.gc.ca/t1/tbl1/en/tv.action?pid=3610043402
% 中国季度growth rate数据 https://data.stats.gov.cn/easyquery.htm?cn=B01&zb=A0103&sj=2024D
% 各国月度growth rate数据 https://www.tradingview.com/support/solutions/43000679714-monthly-gdp-yoy-mgdpyy/?utm_source=chatgpt.com
% 各国家季度gdp数据 https://data.imf.org/regular.aspx?key=63122827
In this section, we apply the MF-SCM method to study the effect of the Tax Cuts and Jobs Act (TCJA) of 2017, enacted by the Trump administration in the U.S. The TCJA aimed to reduce corporate tax rates and provide additional tax cuts for individuals, with the goal of stimulating economic growth. We evaluate the policy, implemented in December 2017, using GDP growth data from several countries. Most countries report GDP growth data on a quarterly basis, while others, such as the United Kingdom, Canada, Russia, and Mexico, report it monthly. In this analysis, we use four high-frequency control units (with monthly data) and ten normal-frequency control units (with quarterly data), selecting the top 15 largest economies by GDP at 2017, with the U.S. as the target group. The pre-treatment period spans 120 months (from December 2008 to December 2017), corresponding to 40 quarterly data points, and the post-treatment period spans 72 months (from January 2018 to December 2023), corresponding to 24 data points. Note that no covariates are included in this experiment, as no low-frequency group is present that would require high-frequency covariates for our algorithm.

\begin{figure}[htbp]
    \centering
    \begin{minipage}[b]{0.7\textwidth}
        \centering
        \includegraphics[width=\textwidth]{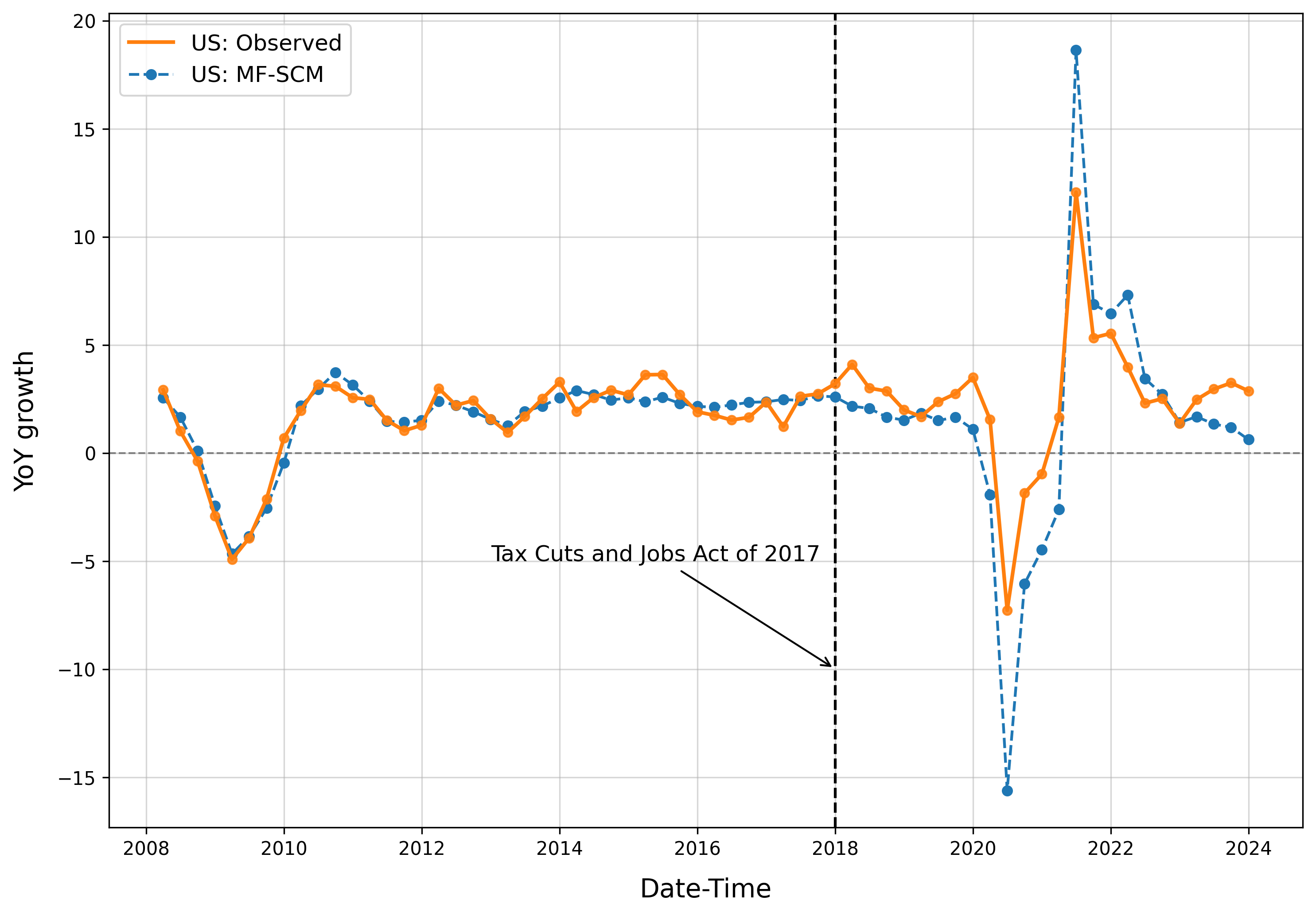}
        \caption{MF-SCM results for the TCJA, showing the observed (green) and synthetic (blue) GDP growth in the U.S. The dashed line indicates the start of the post-treatment period.}
        \label{fig:real7-1-1}
    \end{minipage}
    \\[1ex]
    \begin{minipage}[b]{0.7\textwidth}
        \centering
        \includegraphics[width=\textwidth]{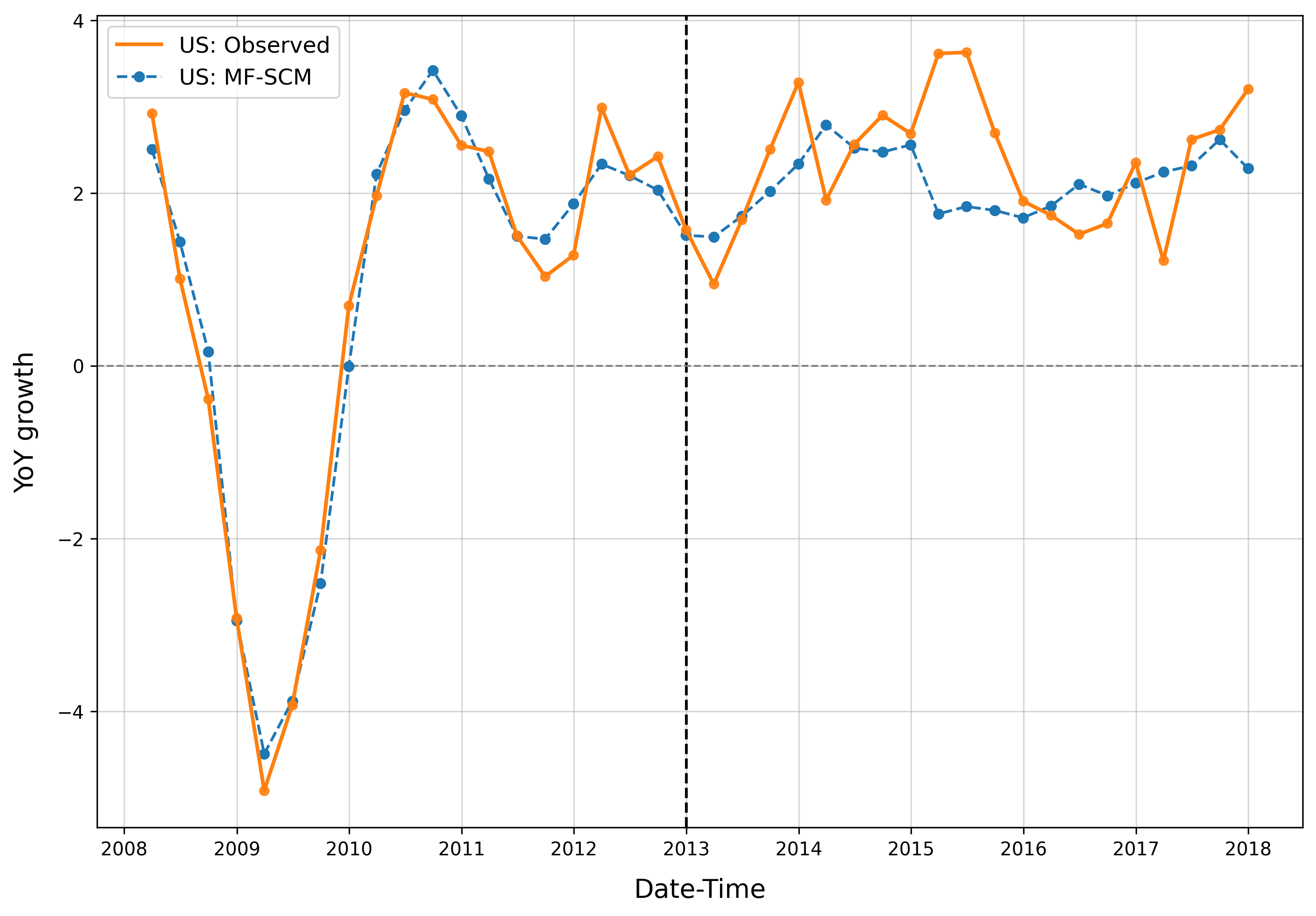}
        \caption{MF-SCM results for the placebo test, showing the observed (green) and synthetic (blue) GDP growth in the U.S. The dashed line indicates the start of the post-treatment period.}
        \label{fig:real7-1-2}
    \end{minipage}
\end{figure}

We first assess the MF-SCM method by evaluating its ability to replicate the actual U.S. GDP growth trajectory during the pre-treatment period. We calculate the MSE between the actual and synthetic U.S. GDP growth during this period, yielding a value of 0.3065. As shown in Figure \ref{fig:real7-1-1}, the synthetic control (blue line) closely follows the actual U.S. GDP growth (orange line) from December 2008 to December 2017, demonstrating a strong pre-treatment fit.

\begin{figure}[htb]
    \centering
    \includegraphics[width=0.7\linewidth]{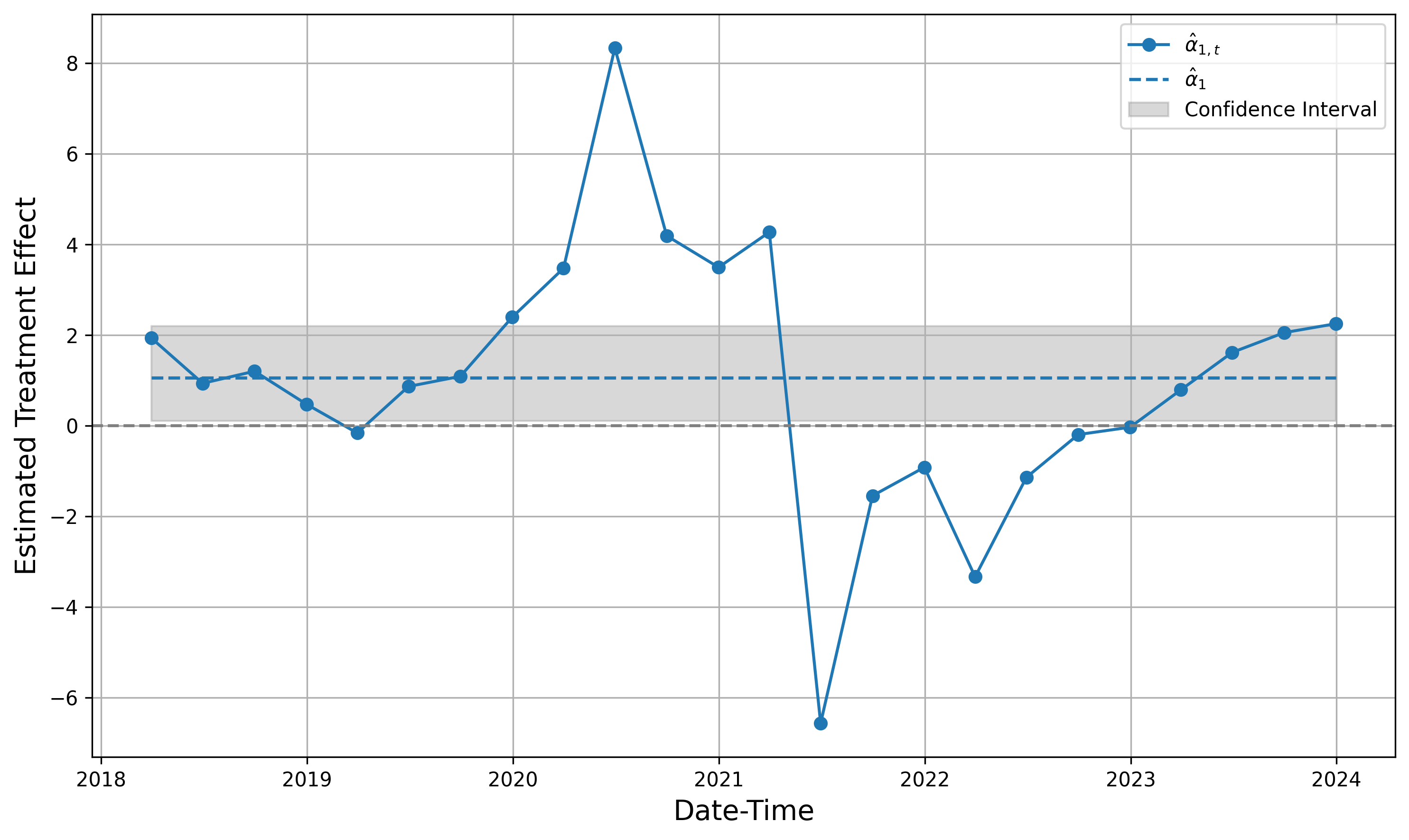}
    \caption{Estimates of treatment effects of TCJA based on MF-SCM.}
    \label{fig:real7-1-3}
\end{figure}

Following the pre-treatment fit, Figure \ref{fig:real7-1-1} illustrates the divergence between the actual and synthetic GDP growth trajectories after December 2017. This divergence suggests that the TCJA had a positive impact on economic performance in the U.S., with GDP growth rising relative to the synthetic counterfactual. The estimated treatment effect $\widehat{\alpha}_{1,t}$ from January 2018 to December 2023 is shown in Figure \ref{fig:real7-1-3}, where the estimated ATE $\widehat{\alpha}_{1}$ is 1.0617 percentage points and the corresponding \(90\%\) confidence interval is \((0.1098, 2.1990)\), so the positive effect is statistically significant at the \(10\%\) level. This positive treatment effect provides evidence that the TCJA contributed to a notable rise in economic activity, reinforcing the conclusion that the policy played a substantial role in stimulating U.S. economic growth during the post-treatment period. We note that the pronounced fluctuations observed beginning in 2020 are largely attributable to the COVID-19 pandemic. After the pandemic, the estimated treatment effects revert to levels similar to those seen prior to the outbreak.

To further confirm the robustness of the observed policy effect, we conduct a placebo test by selecting 2013, a random year before 2017, as a placebo treatment year. In this placebo test, we apply the MF-SCM method as if the TCJA had been implemented in 2013 and generate a synthetic control for that period. We then compare the post-2013 data to the synthetic control to check for any treatment effect. Estimated by the results from 2013 to the end of 2017, the value of ATE is 0.2417, which is much smaller in magnitude than the treatment effect in Figure \ref{fig:real7-1-3}, suggesting that no noticeable effect arises in the placebo setting. See Figure \ref{fig:real7-1-2} for visualization. This suggests that the divergence observed between the actual and synthetic growth post-2017 is indeed due to the TCJA and not to random fluctuations.

Economically, these findings align with the expected mechanisms through which tax cuts could stimulate growth. Lower corporate tax rates reduce the cost of capital, promoting business investment, while individual tax cuts boost consumer spending by increasing disposable income. Both mechanisms work to increase aggregate demand, thereby stimulating overall GDP growth. Our empirical results support this logic, suggesting that the TCJA effectively stimulated economic activity during the post-treatment period.

\subsection{Air Pollution Alerts}\label{sec:7-2}
In this section, we demonstrate the practical applicability of MF-SCM using a real-world dataset from \citet{zheng2024dynamic}, which contains the hourly air pollution data collected in Beijing and neighboring provinces. The dataset includes hourly measurements of PM\(_{2.5}\) concentrations in \(\mu g/m^3\) and meteorological variables, including wind speed, humidity, dew point temperature, and air pressure. These data were collected across 94 monitoring stations, of which 20 are located in Beijing and considered as treated units, while 74 are located in neighboring provinces (Tianjin, Hebei, Shandong, and Shanxi) and used as potential control units. 

We focus on the orange alert which represents a scenario where emission-reduction measures were implemented solely within Beijing, enabling the use of control units to estimate the effects of the alert on PM\(_{2.5}\) concentrations. The analysis focuses on the 48 hours prior to the intervention started at 00:00 on 17 November 2016 and the 24 hours immediately following the start of the alert. We follow the setting used in \citet{zheng2024dynamic}, where the 20 treated units in Beijing are aggregated into a single treated unit by averaging their outcome values and covariates.

To evaluate the proposed MF-SCM in a mixed-frequency environment, we artificially construct mixed-frequency control units by downsampling the original hourly data. This construction is particularly suitable for the present application because the original air pollution data are recorded at specific points in time. That is, each hourly PM$_{2.5}$ observation represents the pollution concentration measured at a particular time point, rather than an aggregate over a longer time interval. Therefore, retaining only one observation every several hours naturally produces lower-frequency point-in-time samples.
This feature corresponds to the special case discussed in the methodological section, where the low-frequency outcome represents a sample taken at a specific point in time. In this case, the unobserved baseline-frequency outcomes for low-frequency control units can be treated as missing observations and recovered through the distributed lag model, rather than being modeled as weighted aggregations of latent baseline-frequency outcomes over a block of periods. Thus, the downsampled hourly pollution data provide a natural and controlled empirical setting for implementing the low-frequency component of MF-SCM.
Specifically, the treated Beijing aggregate is sampled at a 2-hour baseline frequency, while the 74 control units are partitioned into 25 baseline-frequency units observed every 2 hours, 25 low-frequency units observed every 6 hours, and 24 high-frequency units retained at the hourly frequency. 
Starting from the original hourly dataset allows us to construct a full-information benchmark, since all monitoring stations are originally observed at the same high frequency. At the same time, by artificially assigning different sampling frequencies to different control units, we introduce a mixed-frequency structure while keeping the underlying pollution dynamics unchanged. This setup allows us to evaluate the ability of MF-SCM to align mixed-frequency data and estimate treatment effects.

\begin{figure}[htbp]
    \centering
    \includegraphics[width=0.7\textwidth]{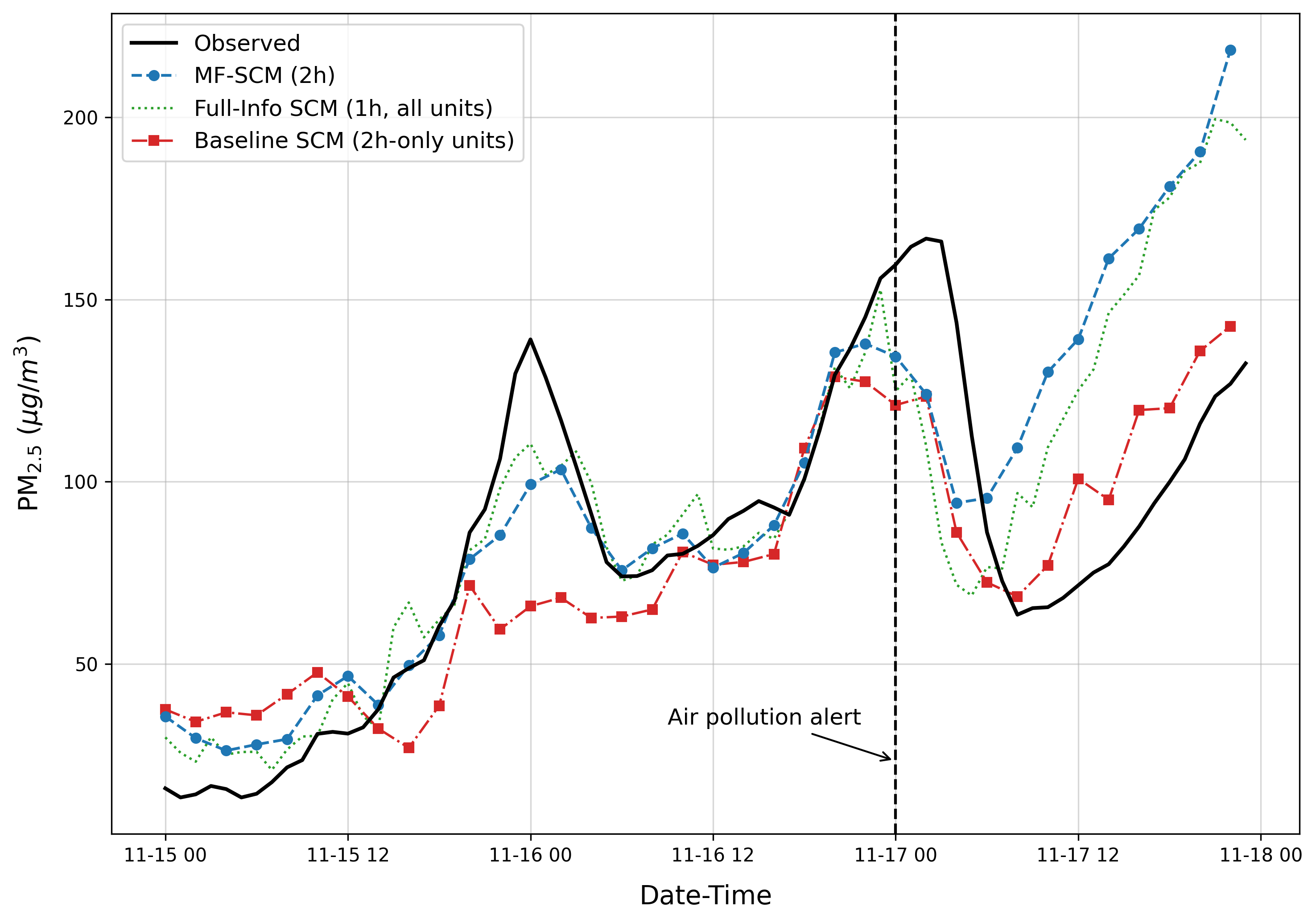}
    \caption{Comparison of observed PM\(_{2.5}\) concentrations and SCM-based counterfactual paths. The dashed line indicates the start of the post-treatment period.}
    \label{fig:results}
\end{figure}

\begin{figure}
    \centering
    \includegraphics[width=0.7\linewidth]{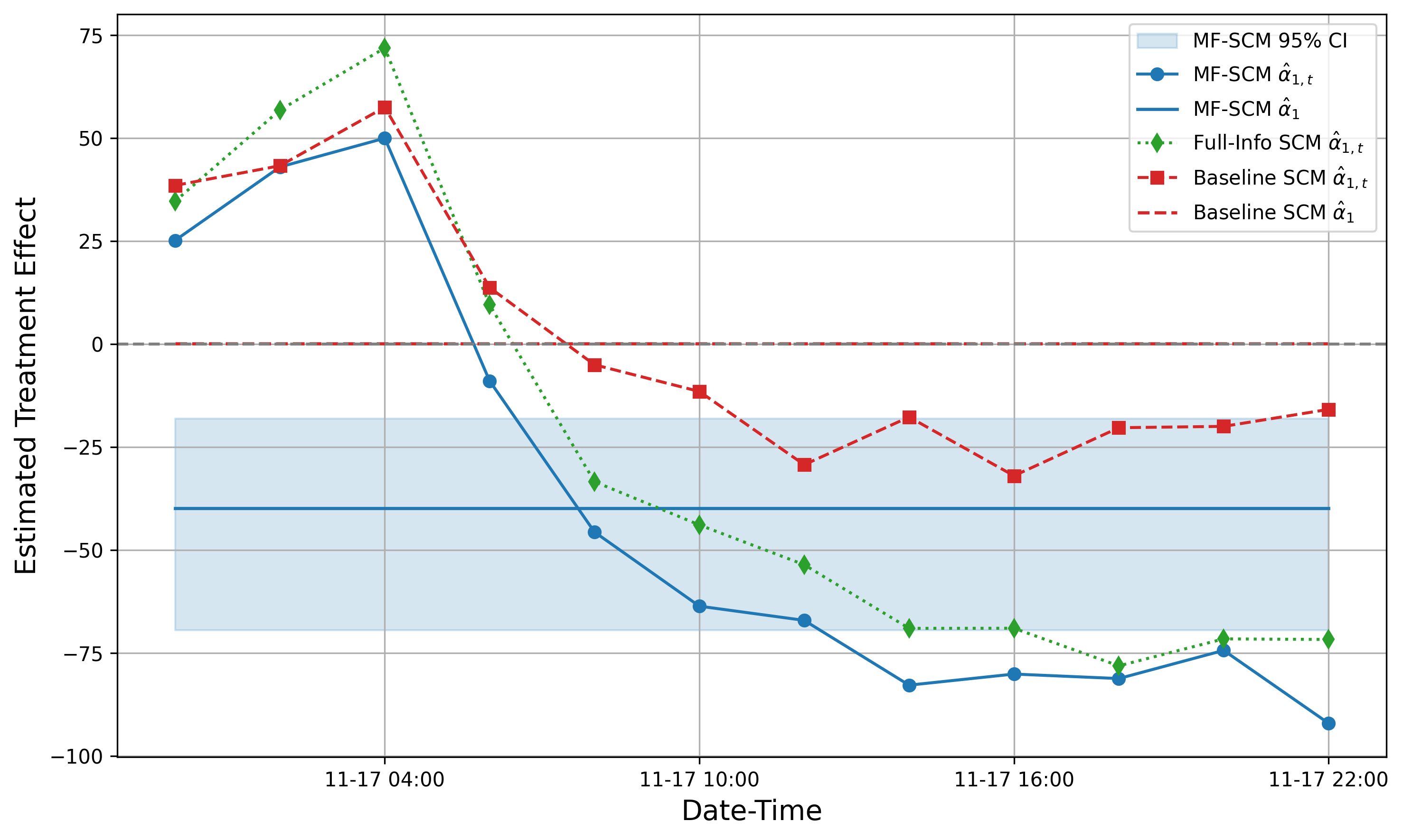}
    \caption{Estimated treatment effects of the alert on PM\(_{2.5}\) concentrations.}
    \label{fig:real7-2-2}
\end{figure}

Figure \ref{fig:results} compares Beijing's observed PM\(_{2.5}\) concentrations with the MF-SCM fit at the 2-hour baseline frequency, the full-information hourly SCM benchmark, and a baseline-frequency-only SCM that uses only the 25 baseline-frequency units. The MF-SCM fit tracks the observed series well before the alert and implies higher counterfactual PM\(_{2.5}\) concentrations after the intervention, so the realized path is consistent with a reduction in pollution once the alert takes effect. In Figure \ref{fig:real7-2-2}, the MF-SCM estimate is \(-40.2445\) with a \(90\%\) confidence interval of \((-70.6683, -18.5985)\), whereas the 2h-only baseline SCM estimate is \(0.1467\) with a \(90\%\) confidence interval of \((-12.7380, 41.1477)\). The corresponding average estimated treatment effect under the full-information hourly SCM benchmark is \(-26.4157\). The estimated treatment path becomes clearly negative several hours after the alert, which is consistent with a delayed response to the emission-reduction measures.

\section{Conclusion}\label{sec:8}
This paper developed a novel mixed-frequency data synthetic control method to address the challenges posed by mixed-frequency data in causal inference. By extending the classical synthetic control methodology, our approach provides a flexible estimation procedure capable of integrating variables observed at different sampling frequency without requiring aggregation or prefiltering. 
We establish the theoretical properties of the MF-SCM estimator, and demonstrate its effectiveness through both simulation studies and empirical applications.

There are several interesting directions for future research. First, the current framework is designed for settings in which the covariates $\mathbf{X}_{j, t}$ are observed at the same or higher frequency than $Y_{1,t}$, which limits its applicability in settings where covariates are available only at lower frequencies. Developing methods that can incorporate lower-frequency covariates would significantly broaden the applicability of MF-SCM. Another important direction is to extend the MF-SCM framework to accommodate more complex data structures, such as missing or censored data, which are common in many real-world datasets and present additional challenges for estimation and inference.

\bibliographystyle{elsarticle-harv} 
\bibliography{refs}

@article{abadie2003economic,
  title={The economic costs of conflict: A case study of the Basque Country},
  author={Abadie, Alberto and Gardeazabal, Javier},
  journal={American Economic Review},
  volume={93},
  number={1},
  pages={113--132},
  year={2003},
  publisher={American Economic Association}
}

@article{abadie2010synthetic,
  title={Synthetic control methods for comparative case studies: Estimating the effect of California's tobacco control program},
  author={Abadie, Alberto and Diamond, Alexis and Hainmueller, Jens},
  journal={Journal of the American Statistical Association},
  volume={105},
  number={490},
  pages={493--505},
  year={2010},
  publisher={Taylor \& Francis}
}

@article{athey2017state,
  title={The state of applied econometrics: Causality and policy evaluation},
  author={Athey, Susan and Imbens, Guido W},
  journal={Journal of Economic Perspectives},
  volume={31},
  number={2},
  pages={3--32},
  year={2017},
  publisher={American Economic Association 2014 Broadway, Suite 305, Nashville, TN 37203-2418}
}

@article{hansen2012jackknife,
  title={Jackknife model averaging},
  author={Hansen, Bruce E and Racine, Jeffrey S},
  journal={Journal of Econometrics},
  volume={167},
  number={1},
  pages={38--46},
  year={2012},
  publisher={Elsevier}
}

@article{botosaru2019role,
  title={On the role of covariates in the synthetic control method},
  author={Botosaru, Irene and Ferman, Bruno},
  journal={The Econometrics Journal},
  volume={22},
  number={2},
  pages={117--130},
  year={2019},
  publisher={Oxford University Press}
}

@article{ferman2021synthetic,
  title={Synthetic controls with imperfect pretreatment fit},
  author={Ferman, Bruno and Pinto, Cristine},
  journal={Quantitative Economics},
  volume={12},
  number={4},
  pages={1197--1221},
  year={2021},
  publisher={Wiley Online Library}
}

@article{ferman2021properties,
  title={On the properties of the synthetic control estimator with many periods and many controls},
  author={Ferman, Bruno},
  journal={Journal of the American Statistical Association},
  volume={116},
  number={536},
  pages={1764--1772},
  year={2021},
  publisher={Taylor \& Francis}
}

@article{zhang2022asymptotic,
  title={Asymptotic Properties of the Synthetic Control Method},
  author={Zhang, Xiaomeng and Wang, Wendun and Zhang, Xinyu},
  journal={arXiv preprint arXiv:2211.12095},
  year={2022}
}

@article{ghysels2004midas,
  title={The MIDAS touch: Mixed data sampling regression models},
  author={Ghysels, Eric and Santa-Clara, Pedro and Valkanov, Rossen},
 journal={Working Papers, UNC and UCLA},
 year={2004}
}

@article{clements2008macroeconomic,
  title={Macroeconomic forecasting with mixed-frequency data: Forecasting output growth in the United States},
  author={Clements, Michael P and Galv{\~a}o, Ana Beatriz},
  journal={Journal of Business \& Economic Statistics},
  volume={26},
  number={4},
  pages={546--554},
  year={2008},
  publisher={Taylor \& Francis}
}

@article{ghysels2005there,
  title={There is a risk-return trade-off after all},
  author={Ghysels, Eric and Santa-Clara, Pedro and Valkanov, Rossen},
  journal={Journal of Financial Economics},
  volume={76},
  number={3},
  pages={509--548},
  year={2005},
  publisher={Elsevier}
}

@book{griffiths1985theory,
  title={The Theory and Practice of Econometrics},
  author={Griffiths, William E and Judge, George G and Hill, R Carter and L{\"u}tkepohl, Helmut and Lee, Tsoung-Chao},
  year={1985},
  publisher={Wiley}
}

@article{zheng2024dynamic,
  title={Dynamic synthetic control method for evaluating treatment effects in auto-regressive processes},
  author={Zheng, Xiangyu and Chen, Song Xi},
  journal={Journal of the Royal Statistical Society Series B: Statistical Methodology},
  volume={86},
  number={1},
  pages={155--176},
  year={2024},
  publisher={Oxford University Press US}
}

@article{hsiao2012panel,
  title={A panel data approach for program evaluation: measuring the benefits of political and economic integration of Hong Kong with mainland China},
  author={Hsiao, Cheng and Steve Ching, H and Ki Wan, Shui},
  journal={Journal of Applied Econometrics},
  volume={27},
  number={5},
  pages={705--740},
  year={2012},
  publisher={Wiley Online Library}
}

@book{politis1999subsampling,
  title={Subsampling in the IID Case},
  author={Politis, Dimitris N and Romano, Joseph P and Wolf, Michael and Politis, Dimitris N and Romano, Joseph P and Wolf, Michael},
  year={1999},
  publisher={Springer}
}

@article{li2017estimation,
  title={Estimation of average treatment effects with panel data: Asymptotic theory and implementation},
  author={Li, Kathleen T and Bell, David R},
  journal={Journal of Econometrics},
  volume={197},
  number={1},
  pages={65--75},
  year={2017},
  publisher={Elsevier}
}

@book{zarantonello1971projections,
  title={Projections on Convex Sets in Hilbert Space and Spectral Theory: Part I. Projections on Convex Sets: Part II. Spectral Theory, in Contributions to Nonlinear Functional Analysis},
  author={Zarantonello, Eduardo H},
  pages={237--424},
  year={1971},
  publisher={Elsevier}
}

@article{fang2019inference,
  title={Inference on directionally differentiable functions},
  author={Fang, Zheng and Santos, Andres},
  journal={The Review of Economic Studies},
  volume={86},
  number={1},
  pages={377--412},
  year={2019},
  publisher={Oxford University Press}
}

@article{li2020statistical,
  title={Statistical inference for average treatment effects estimated by synthetic control methods},
  author={Li, Kathleen T},
  journal={Journal of the American Statistical Association},
  volume={115},
  number={532},
  pages={2068--2083},
  year={2020},
  publisher={Taylor \& Francis}
}

@article{andrews2000inconsistency,
  title={Inconsistency of the bootstrap when a parameter is on the boundary of the parameter space},
  author={Andrews, Donald WK},
  journal={Econometrica},
  volume={68},
  pages={399--405},
  year={2000},
  publisher={JSTOR}
}

@article{amemiya1967comparative,
  title={A comparative study of alternative estimators in a distributed lag model},
  author={Amemiya, Takeshi and Fuller, Wayne A},
  journal={Econometrica},
  volume={35},
  pages={509--529},
  year={1967},
  publisher={JSTOR}
}

@article{almon1965distributed,
  title={The distributed lag between capital appropriations and expenditures},
  author={Almon, Shirley},
  journal={Econometrica},
  volume={33},
  pages={178--196},
  year={1965},
  publisher={JSTOR}
}

@inproceedings{sun2024temporal,
  title={Temporal aggregation for the synthetic control method},
  author={Sun, Liyang and Ben-Michael, Eli and Feller, Avi},
  booktitle={AEA Papers and Proceedings},
  volume={114},
  pages={614--617},
  year={2024},
  organization={American Economic Association 2014 Broadway, Suite 305, Nashville, TN 37203}
}

@article{abadie2022synthetic,
  title={Synthetic controls in action},
  author={Abadie, Alberto and Vives-i-Bastida, Jaume},
  journal={arXiv preprint arXiv:2203.06279},
  year={2022}
}

@article{zhang2024asymptotic,
  title={Asymptotic Properties of the Distributional Synthetic Controls},
  author={Zhang, Lu and Zhang, Xiaomeng and Zhang, Xinyu},
  journal={arXiv preprint arXiv:2405.00953},
  year={2024}
}

@article{zhu2023synthetic,
  title={Synthetic Regressing Control Method},
  author={Zhu, Rong JB},
  journal={arXiv preprint arXiv:2306.02584},
  year={2023}
}

@article{babii2022machine,
  title={Machine learning time series regressions with an application to nowcasting},
  author={Babii, Andrii and Ghysels, Eric and Striaukas, Jonas},
  journal={Journal of Business \& Economic Statistics},
  volume={40},
  number={3},
  pages={1094--1106},
  year={2022},
  publisher={Taylor \& Francis}
}

@article{marsilli2014variable,
  title={Variable selection in predictive MIDAS models},
  author={Marsilli, Cl{\'e}ment},
  year={2014},
  publisher={Banque de France working paper}
}

@article{foroni2013survey,
  title={A survey of econometric methods for mixed-frequency data},
  author={Foroni, Claudia and Marcellino, Massimiliano Giuseppe},
  journal={Available at SSRN 2268912},
  year={2013}
}

@article{andrews1999estimation,
  title={Estimation when a parameter is on a boundary},
  author={Andrews, Donald WK},
  journal={Econometrica},
  volume={67},
  number={6},
  pages={1341--1383},
  year={1999},
  publisher={Wiley Online Library}
}

@article{geyer1994asymptotics,
  title={On the asymptotics of constrained M-estimation},
  author={Geyer, Charles J},
  journal={The Annals of statistics},
  pages={1993--2010},
  year={1994},
  publisher={JSTOR}
}

\end{document}